\documentstyle [10pt,amsfonts] {article}
\input epsf

\newcommand{\beq}{\begin{equation}}
\newcommand{\eeq}{\end{equation}}
\newcommand{\beqs}{\begin{eqnarray}}
\newcommand{\eeqs}{\end{eqnarray}}

\catcode`@=11
\@addtoreset{equation}{section}
\@addtoreset{equation}{subsection}
\def\theequation{\ifnum\value{section}=0 \arabic{equation}\ignorespaces
\else \ifnum\value{section}=-1 A.\arabic{equation}\ignorespaces
\else \ifnum\value{subsection}=0 \thesection.\arabic{equation}\ignorespaces
\else \thesection.\arabic{subsection}.\arabic{equation}\ignorespaces
                           \fi
                      \fi
                 \fi}
\catcode`@=12

\begin{document}

\def\thefootnote{\fnsymbol{footnote}}

\baselineskip 5.0mm

\vspace{4mm}

\begin{center}

{\Large \bf Exact Potts Model Partition Function on Strips of the 
Triangular Lattice}

\vspace{8mm}

\setcounter{footnote}{0}
Shu-Chiuan Chang$^{(a)}$\footnote{email: shu-chiuan.chang@sunysb.edu} and
\setcounter{footnote}{6}
Robert Shrock$^{(a,b)}$\footnote{(a): permanent address;
email: robert.shrock@sunysb.edu}

\vspace{6mm}

(a) \ C. N. Yang Institute for Theoretical Physics  \\
State University of New York       \\
Stony Brook, N. Y. 11794-3840  \\

(b) \ Physics Department \\
Brookhaven National Laboratory \\
Upton, NY  11973

\vspace{10mm}

{\bf Abstract}
\end{center}

In this paper we present exact calculations of the partition function
$Z$ of the $q$-state Potts model and its generalization to real $q$, 
for arbitrary temperature on $n$-vertex strip graphs, of width
$L_y=2$ and arbitrary length, of the triangular lattice with free, cyclic, and
M\"obius longitudinal boundary conditions. These partition functions are
equivalent to Tutte/Whitney polynomials for these graphs.  The free energy is
calculated exactly for the infinite-length limit of the graphs, and the
thermodynamics is discussed.  Considering the full generalization to arbitrary
complex $q$ and temperature, we determine the singular locus ${\cal B}$ in the
corresponding ${\mathbb C}^2$ space, arising as the accumulation set of
partition function zeros as $n \to \infty$.  In particular, we study the
connection with the $T=0$ limit of the Potts antiferromagnet where ${\cal B}$
reduces to the accumulation set of chromatic zeros.  Comparisons are made with
our previous exact calculation of Potts model partition functions for the
corresponding strips of the square lattice.  Our present calculations yield, as
special cases, several quantities of graph-theoretic interest.

\vspace{16mm}

\pagestyle{empty}
\newpage

\pagestyle{plain}
\pagenumbering{arabic}
\renewcommand{\thefootnote}{\arabic{footnote}}
\setcounter{footnote}{0}

\section{Introduction}

The $q$-state Potts model has served as a valuable model for the study of phase
transitions and critical phenomena \cite{potts,wurev}.  On a lattice, or, more
generally, on a (connected) graph $G$, at temperature $T$, this model is 
defined by the partition function
\beq
Z(G,q,v) = \sum_{ \{ \sigma_n \} } e^{-\beta {\cal H}}
\label{zfun}
\eeq
with the (zero-field) Hamiltonian
\beq
{\cal H} = -J \sum_{\langle i j \rangle} \delta_{\sigma_i \sigma_j}
\label{ham}
\eeq
where $\sigma_i=1,...,q$ are the spin variables on each vertex $i \in G$;
$\beta = (k_BT)^{-1}$; and $\langle i j \rangle$ denotes pairs of adjacent
vertices.  The graph $G=G(V,E)$ is defined by its vertex set $V$ and its edge
set $E$; we denote the number of vertices of $G$ as $n=n(G)=|V|$ and the
number of edges of $G$ as $e(G)=|E|$.  We use the notation
\beq
K = \beta J \ , \quad a = u^{-1} = e^K \ , \quad v = a-1
\label{kdef}
\eeq
so that the physical ranges are (i) $a \ge 1$, i.e., $v \ge 0$ corresponding to
$\infty \ge T \ge 0$ for the Potts ferromagnet, and (ii) $0 \le a \le 1$,
i.e., $-1 \le v \le 0$, corresponding to $0 \le T \le \infty$ for the Potts
antiferromagnet.  One defines the (reduced) free energy per site $f=-\beta F$,
where $F$ is the actual free energy, via
\beq
f(\{G\},q,v) = \lim_{n \to \infty} \ln [ Z(G,q,v)^{1/n}]  \ .
\label{ef}
\eeq
where we use the symbol $\{G\}$ to denote $\lim_{n \to \infty}G$ for a given
family of graphs.

Let $G^\prime=(V,E^\prime)$ be a spanning subgraph of $G$, i.e. a subgraph
having the same vertex set $V$ and an edge set $E^\prime \subseteq E$. Then
$Z(G,q,v)$ can be written as the sum \cite{birk}-\cite{kf}
\beqs
Z(G,q,v) & = & \sum_{G^\prime \subseteq G} q^{k(G^\prime)}v^{e(G^\prime)}
\label{cluster} \cr\cr\cr
& = & \sum_{r=k(G)}^{n(G)}\sum_{s=0}^{e(G)}z_{rs} q^r v^s
\label{zpol}
\eeqs
where $k(G^\prime)$ denotes the number of connected components of $G^\prime$
and $z_{rs} \ge 0$.  Since we only consider connected graphs $G$, we have
$k(G)=1$. The formula (\ref{cluster}) enables one to generalize $q$
from ${\mathbb Z}_+$ to ${\mathbb R}_+$ (keeping $v$ in its physical range).
This generalization is sometimes denoted the random cluster model \cite{kf};
here we shall use the term ``Potts model'' to include both positive integral
$q$ as in the original formulation in eqs. (\ref{zfun}) and (\ref{ham}), and
the generalization to real (or complex) $q$, via eq. (\ref{cluster}).  The
formula (\ref{cluster}) shows that $Z(G,q,v)$ is a polynomial in $q$ and $v$
(equivalently, $a$) with maximum and minimum degrees indicated in
eq. (\ref{zpol}).  The Potts model partition function on
a graph $G$ is essentially equivalent to the Tutte polynomial
\cite{tutte1}-\cite{tutte3} and Whitney rank polynomial \cite{whit},
\cite{wurev}, \cite{bbook}-\cite{boll} for this graph, as discussed in the
appendix.

\vspace{6mm}

In this paper we shall present exact calculations of the Potts model partition
function for strips of the triangular lattice of width $L_y=2$ vertices (or
equivalently edges) and arbitrary length $L_x$ with various boundary
conditions.  This is a natural continuation of a previous study by one of us of
the Potts model model on the analogous strips of the square lattice
\cite{bcc,a}, and the reader is referred to that paper for background and
further references.  We envision the strip of the triangular lattice as being
formed by starting with a ladder graph, i.e. $2 \times L_x$ strip of the square
lattice, and then adding edges joining, say, the lower left and upper right
vertices of each square.  The longitudinal (transverse) direction is taken as
the horizontal, $x$ (vertical, $y$) direction.  We use free transverse boundary
conditions and consider free, periodic (= cyclic), and M\"obius longitudinal
boundary conditions.  These families of graphs are denoted, respectively, as
$S_{m}$ (for open strip), $L_m$ (for ladder), and $ML_m$ (for M\"obius ladder),
where $L_x=m+1$ (edges) for $S_m$ (following our labelling convention in
\cite{strip}) and $L_x=m$ for $L_m$ and $ML_m$.  One has $n(S_m)=2(m+2)$ and
$n(L_m)=n(ML_m)=2m$.  Each vertex on the cyclic strip $L_m$ has degree
(coordination number) $\Delta=4$; this is also true of the interior vertices on
the open strip $S_m$.  The M\"obius strip can involve a seam, as discussed in
\cite{t}.

The motivations for these exact calculations of Potts model partition functions
for strips of various lattices were discussed in \cite{a}.  Clearly, new exact
calculations of Potts model partition functions are of value in their own
right.  In addition, it was shown \cite{a} that these calculations can give
insight into the complex-temperature phase diagram of the 2D Potts model on the
given lattice.  This is useful, since the 2D Potts model has never been solved
except in the $q=2$ Ising case. Furthermore, with these exact results one can
study both the $T=0$ critical point of the $q$-state Potts ferromagnet and, for
certain $q$ values ($q=2$ for the square strip; $q=2,3$ for the present strip
of the triangular lattice) the $T=0$ critical point of the Potts
antiferromagnet.  In addition, via the correspondence with the Tutte
polynomial, our calculations yield several quantities of relevance to
mathematical graph theory.

Various special cases of the Potts model partition function are of interest.
One special case is the zero-temperature limit of the Potts antiferromagnet
(AF). For sufficiently large $q$, on a given lattice or graph $G$, this 
exhibits nonzero ground state entropy (without frustration).
This is equivalent to a ground
state degeneracy per site (vertex), $W > 1$, since $S_0 = k_B \ln W$.  The
$T=0$ (i.e., $v=-1$) partition function of the above-mentioned $q$-state Potts
antiferromagnet (PAF) on $G$ satisfies
 \beq Z(G,q,-1)=P(G,q)
\label{zp}
\eeq
where $P(G,q)$ is the chromatic polynomial (in $q$) expressing the number
of ways of coloring the vertices of the graph $G$ with $q$ colors such that no
two adjacent vertices have the same color \cite{birk,bbook,rrev,rtrev}. The 
minimum number of colors necessary for this coloring is the chromatic number
of $G$, denoted $\chi(G)$.  We have 
\beq 
W(\{G\},q)= \lim_{n \to \infty}P(G,q)^{1/n}
\label{w}
\eeq

Using the formula (\ref{cluster}) for $Z(G,q,v)$, one can generalize $q$ from
${\mathbb Z}_+$ not just to ${\mathbb R}_+$ but to ${\mathbb C}$ and $a$ from
its physical ferromagnetic and antiferromagnetic ranges $1 \le a \le \infty$
and $0 \le a \le 1$ to $a \in {\mathbb C}$.  A subset of the zeros of $Z$ in
the two-complex dimensional space ${\mathbb C}^2$ defined by the pair of
variables $(q,a)$ can form an accumulation set in the $n \to \infty$ limit,
denoted ${\cal B}$, which is the continuous locus of points where the free
energy is nonanalytic.  This locus is determined as the solution to a certain
$\{G\}$-dependent equation \cite{bcc,a}.  For a given value of $a$, one can
consider this locus in the $q$ plane, and we denote it as ${\cal
B}_q(\{G\},a)$.  In the special case $a=0$ (i.e., $v=-1$) where the partition
function is equal to the chromatic polynomial, the zeros in $q$ are the
chromatic zeros, and ${\cal B}_q(\{G\},a=0)$ is their continuous accumulation
set in the $n \to \infty$ limit \cite{rrev}-\cite{t}.  In a series of papers
we have given exact calculations of the chromatic polynomials and nonanalytic
loci ${\cal B}_q$ for various families of graphs (for further references on
this $a=0$ special case, see \cite{a}).  With the exact Potts partition
function for arbitrary temperature, one can study ${\cal B}_q$ for $a \ne 0$
and, for a given value of $q$, one can study the continuous accumulation set of
the zeros of $Z(G,q,v)$ in the $a$ plane; this will be denoted ${\cal
B}_a(\{G\},q)$.  It will often be convenient to consider the equivalent locus
in the $u=1/a$ plane, namely ${\cal B}_u(\{G\},q)$.  We shall sometimes write
${\cal B}_q(\{G\},a)$ simply as ${\cal B}_q$ when $\{G\}$ and $a$ are clear
from the context, and similarly with ${\cal B}_{a}$ and ${\cal B}_{u}$.  One
gains a unified understanding of the separate loci ${\cal B}_q(\{G\})$ for
fixed $a$ and ${\cal B}_a(\{G\})$ for fixed $q$ by relating these as different
slices of the locus ${\cal B}$ in the ${\mathbb C}^2$ space defined by $(q,a)$
as we shall do here.

Following the notation in \cite{w} and our other earlier works on ${\cal
B}_q(\{G\})$ for $a=0$, we denote the maximal region in the complex $q$ plane
to which one can analytically continue the function $W(\{G\},q)$ from physical
values where there is nonzero ground state entropy as $R_1$ .  The maximal
value of $q$ where ${\cal B}_q$ intersects the (positive) real axis was
labelled $q_c(\{G\})$.  Thus, region $R_1$ includes the positive real axis for
$q > q_c(\{G\})$.  Correspondingly, in our works on complex-temperature
properties of spin models, we had labelled the complex-temperature extension
(CTE) of the physical paramagnetic phase as (CTE)PM, which will simply be
denoted PM here, the extension being understood, and similarly with
ferromagnetic (FM) and antiferromagnetic (AFM); other complex-temperature
phases, having no overlap with any physical phase, were denoted $O_j$ (for
``other''), with $j$ indexing the particular phase \cite{chisq}.  Here we shall
continue to use this notation for the respective slices of ${\cal B}$ in the
$q$ and $a$ or $u$ planes.

We record some special values of $Z(G,q,v)$ below, beginning with the $q=0$
special case
\beq
Z(G,0,v)=0
\label{zq0}
\eeq
which implies that $Z(G,q,v)$ has an overall factor of $q$. In general
(and for all the graphs considered here), this is the only overall factor that
it has.  We also have
\beq
Z(G,1,v)=\sum_{G^\prime \subseteq G} v^{e(G^\prime)} = a^{e(G)} \ .
\label{zq1}
\eeq
For temperature $T=\infty$, i.e., $v=0$, 
\beq
Z(G,q,0)=q^{n(G)} \ .
\label{za1}
\eeq
Further, 
\beq
Z(G,q,-1)=P(G,q)=\biggl [ \prod_{s=0}^{\chi(G)-1}(q-s) \biggr ] U(G,q)
\label{pchi}
\eeq
where $U(G,q)$ is a polynomial in $q$ of degree $n(G)-\chi(G)$. 

We note some values of chromatic numbers for the strip graphs considered here:
\beq
\chi(S_m)=3
\label{chism}
\eeq
\beq
\chi(L_m) = \cases{ 3 & if $m=0$ \
mod 3 \cr 4 & if $m=1$ \ or $m=2$ mod 3 \cr }
\label{chitly2}
\eeq
and
\beq
\chi(ML_m)=4 \ . 
\label{chittly2}
\eeq

Hence, for $q=1,2,3$ we have
\beq
Z(G,q,-1)=P(G,q)=0 \quad {\rm for} \quad G=S_m, \ L_m, \ ML_m \quad 
{\rm and} \quad q=1,2
\label{pq12}
\eeq
\beq
Z(G,3,-1)=P(G,3)=3! \quad {\rm for} \quad G=S_m \ \  {\rm or} \ \ 
G=L_{m=0 \ mod \ 3}
\label{pq3}
\eeq
where $G=L_{m=0 \ mod \ 3}$ means that this applies to $L_m$ for $m=0$ mod 3. 
For the graphs $G=S_m,L_m$, and $ML_m$, (i) the result (\ref{pq12}) implies 
that for $q=2$, $Z(G,2,v)$ contains at least one power of the factor 
$(v+1)=a$; for $q=1$, one already knows the form of $Z(G,1,v)$ from 
(\ref{zq1}); (ii)  the result (\ref{pq3}) implies that for the cases where 
$\chi(G)=4$, $Z(G,3,v)$ contains at least one power of $(v+1)$ as a factor. 

Another basic property, evident from eq. (\ref{cluster}), is that (i) the zeros
of $Z(G,q,v)$ in $q$ for real $v$ and hence also the continuous accumulation
set ${\cal B}_q$ are invariant under the complex conjugation $q \to q^*$; (ii)
the zeros of $Z(G,q,v)$ in $v$ or equivalently $a$ for real $q$ and hence also
the continuous accumulation set ${\cal B}_a$ are invariant under the complex
conjugation $a \to a^*$.

Just as the importance of noncommutative limits was shown in (eq. (1.9) of)
\cite{w} on chromatic polynomials, so also one encounters an analogous
noncommutativity here for the general partition function (\ref{cluster}) of
the Potts model for nonintegral $q$: at certain special points $q_s$
(typically $q_s=0,1...,\chi(G)$) one has
\beq 
\lim_{n \to \infty} \lim_{q
\to q_s} Z(G,q,v)^{1/n} \ne \lim_{q \to q_s} \lim_{n \to \infty} Z(G,q,v)^{1/n}
\ .
\label{fnoncomm}
\eeq

Because of
this noncommutativity, the formal definition (\ref{ef}) is, in general,
insufficient to define the free energy $f$ at these special points $q_s$; it is
necessary to specify the order of the limits that one uses in eq.
(\ref{fnoncomm}).  We denote the two
definitions using different orders of limits as $f_{qn}$ and $f_{nq}$:
\beq
f_{nq}(\{G\},q,v) = \lim_{n \to \infty} \lim_{q \to q_s} n^{-1} \ln Z(G,q,v)
\label{fnq}
\eeq
\beq
f_{qn}(\{G\},q,v) = \lim_{q \to q_s} \lim_{n \to \infty} n^{-1} \ln Z(G,q,v) \
.
\label{fqn}
\eeq
In Ref. \cite{w} and our subsequent works on chromatic polynomials and the
above-mentioned zero-temperature antiferromagnetic limit, it was convenient to
use the ordering $W(\{G\},q_s) = \lim_{q \to q_s} \lim_{n \to \infty}
P(G,q)^{1/n}$ since this avoids certain discontinuities in $W$ that would be
present with the opposite order of limits.  In the present work on the full
temperature-dependent Potts model partition function, we shall
consider both orders of limits and comment on the differences where
appropriate.  Of course in discussions of the usual $q$-state Potts model (with
positive integer $q$), one automatically uses the definition in eq.
(\ref{zfun}) with (\ref{ham}) and no issue of orders of limits arises, as it
does in the Potts model with real $q$.

As a consequence of the noncommutativity (\ref{fnoncomm}), it follows that for
the special set of points $q=q_s$ one must distinguish between (i) $({\cal
B}_a(\{G\},q_s))_{nq}$, the continuous accumulation set of the zeros of
$Z(G,q,v)$ obtained by first setting $q=q_s$ and then taking $n \to \infty$,
and (ii) $({\cal B}_a(\{G\},q_s))_{qn}$, the continuous accumulation set of the
zeros of $Z(G,q,v)$ obtained by first taking $n \to \infty$, and then taking $q
\to q_s$.  For these special points,
\beq
({\cal B}_a(\{G\},q_s))_{nq} \ne ({\cal B}_a(\{G\},q_s))_{qn} \ .
\label{bnoncomm}
\eeq

 From eq. (\ref{zq0}), it follows that for any $G$,
\beq
\exp(f_{nq})=0 \quad {\rm for} \quad q=0
\label{fnqq0}
\eeq
and thus
\beq
({\cal B}_a)_{nq} = \emptyset \quad {\rm for} \quad q=0 \ .
\label{bnq0}
\eeq 
However, for many families of graphs, including the circuit graph $C_n$,
and cyclic and M\"obius strips of the square or triangular lattice, if we take
$n \to \infty$ first and then $q \to 0$, we find that $({\cal B}_u)_{qn}$ is
nontrivial.  Similarly, from (\ref{zq1}) we have, for any $G$, 
\beq 
({\cal B}_a)_{nq} = \emptyset \quad {\rm for} \quad q=1
\label{bnq1}
\eeq
since all of the zeros of $Z$ occur at the single discrete point $a=0$ (and in
the case of a graph $G$ with no edges, $Z=1$ with no zeros).  However, as the
simple case of the circuit graph shows \cite{a}, $({\cal B}_u)_{qn}$ is, in 
general, nontrivial.

As derived in \cite{a}, a general form for the Potts model partition function
for the strip graphs considered here, or more generally, for recursively
defined families of graphs comprised of $m$ repeated subunits (e.g. the columns
of squares of height $L_y$ vertices that are repeated $L_x$ times to form an
$L_x \times L_y$ strip of a regular lattice with some specified boundary
conditions), is 
\beq 
Z(G,q,v) = \sum_{j=1}^{N_\lambda} c_{G,j}
(\lambda_{G,j}(q,v))^m
\label{zgsum}
\eeq
where $N_\lambda$ depends on $G$.

The Potts ferromagnet has a zero-temperature phase transition in the $L_x \to
\infty$ limit of the strip graphs considered here, and this has the consequence
that for cyclic and M\"obius boundary conditions, ${\cal B}$ passes through the
$T=0$ point $u=0$.  It follows that ${\cal B}$ is noncompact in the $a$ plane.
Hence, it is usually more convenient to study the slice of ${\cal B}$ in the
$u=1/a$ plane rather than the $a$ plane.  Since $a \to
\infty$ as $T \to 0$ and $Z$ diverges like $a^{e(G)}$ in this limit, we shall
use the reduced partition function $Z_r$ defined by
\beq
Z_r(G,q,v)=a^{-e(G)}Z(G,q,v)=u^{e(G)}Z(G,q,v)
\label{zr}
\eeq
which has the finite limit $Z_r \to 1$ as $T \to 0$.  For a general strip
graph $(G_s)_m$ of type $G_s$ and length $L_x=m$, we can write
\beqs
Z_r((G_s)_m,q,v) & = & u^{e((G_s)_m)}\sum_{j=1}^{N_\lambda} c_{G_s,j}
(\lambda_{G_s,j})^m \equiv \sum_{j=1}^{N_\lambda} c_{G_s,j}
(\lambda_{G_s,j,u})^m
\label{zu}
\eeqs
with
\beq
\lambda_{G_s,j,u}=u^{e((G_s)_m)/m}\lambda_{G_s,j} \ .
\label{lamu}
\eeq
For example, for the $L_y=2$ strips of the triangular lattice of
interest here, with free transverse boundary conditions and either free or
periodic longitudinal boundary conditions, we have $e(S_m)=4m+5$ and 
$e(L_m)=4m$.  (In the case of the cyclic strip with $m=2$, $L_m$ degenerates in
the sense that it contains multiple edges connecting some pairs of vertices, 
and similarly, for $m=1$, $L_m$ contains both multiple edges and loops; the 
above formula applies to these cases if one takes care to count these multiple
edges and loops, as is discussed further below.) 

\section{Strip of Triangular Lattice with Free Longitudinal Boundary 
Conditions}

In this section we present the Potts model partition function
$Z(S_m,q,v)$ for the $L_y=2$ strip of the triangular 
lattice $S_m$ with arbitrary
length $L_x=m+1$ (i.e., containing $m+1$ squares) and free transverse and
longitudinal boundary conditions.
One convenient way to express the results is in terms of a generating
function:
\beq
\Gamma(S,q,v,z) = \sum_{m=0}^\infty Z(S_m,q,v)z^m \ .
\label{gammazfbc}
\eeq
We have calculated this generating function using the deletion-contraction
theorem for the corresponding Tutte polynomial $T(S_m,x,y)$ and then
expressing the result in terms of the variables $q$ and $v$.  We find
\beq
\Gamma(S,q,v,z) = \frac{ {\cal N}(S,q,v,z)}{{\cal D}(S,q,v,z)}
\label{gammazcalc}
\eeq
where
\beq
{\cal N}(S,q,v,z)=A_{S,0}+A_{S,1}z
\label{numgamma}
\eeq
with
\beq
A_{S,0}=q(v^5+5v^4+8v^3+2v^3q+10v^2q+5vq^2+q^3)
\label{as0}
\eeq
\beq
A_{S,1}=-q(v+1)^2(v+q)^3v^2
\label{as1}
\eeq
and
\beq
{\cal D}(S,q,v,z) = 1-(v^4+4v^3+7v^2+4qv+q^2)z+(v+1)^2(v+q)^2v^2z^2 \ .
\label{dengamma}
\eeq
(The generating function for the Tutte polynomial
$T(S_m,x,y)$ is given in the appendix.) Writing
\beq
{\cal D}(S,q,v,z) = \prod_{j=1}^2 (1-\lambda_{S,j}z)
\label{ds}
\eeq
we have
\beq
\lambda_{S,(1,2)} = \frac{1}{2} \biggl [ T_{S12} \pm (3v+v^2+q)\sqrt{R_{S12}}
 \ \biggr ]
\label{lams}
\eeq
where
\beq
T_{S12}=v^4+4v^3+7v^2+4qv+q^2
\label{t12}
\eeq
and
\beq
R_{S12} = q^2+2qv-2qv^2+5v^2+2v^3+v^4 \ . 
\label{rs12}
\eeq

Ref. \cite{hs} presented a formula to obtain the chromatic polynomial
for a recursive family of graphs in the form of sums of powers of
$\lambda_j$'s starting from the generating function, and the
generalization of this to the full Potts model partition function was
given in \cite{a}.  Using this, we have
\beq
Z(S_m,q,v) = \frac{(A_{S,0}\lambda_{S,1} + A_{S,1})}
{(\lambda_{S,1}-\lambda_{S,2})}\lambda_{S,1}^m +
 \frac{(A_{S,0} \lambda_{S,2} + A_{S,1})}
{(\lambda_{S,2}-\lambda_{S,1})}\lambda_{S,2}^m
\label{pgsumkmax2}
\eeq 
(which is symmetric under $\lambda_{S,1} \leftrightarrow \lambda_{S,2}$).
Although both the $\lambda_{S,j}$'s and the coefficient functions involve the
square root $\sqrt{R_{S12}}$ and are not polynomials in $q$ and $v$, the
theorem on symmetric functions of the roots of an algebraic equation
\cite{uspensky} guarantees that $Z(S_m,q,v)$ is a polynomial in $q$ and $v$ (as
it must be by (\ref{cluster}) since the coefficients of the powers of $z$ in
the equation (\ref{dengamma}) defining these $\lambda_{S,j}$'s are polynomials
in these variables $q$ and $v$.

As will be shown below, for $q \ge 2$ (and for $0 \le q < 1$, with the 
$f_{qn}$ and $({\cal B}_u)_{qn}$ definitions) the singular locus 
${\cal B}_u$ for this strip consists of arcs that
do not separate the $u$ plane into different regions, so that the PM phase
and its complex-temperature extension occupy all of this plane, except for
these arcs.  For these ranges of $q$, the (reduced) free energy is given by
\beq
f = \frac{1}{2}\ln \lambda_{S,1} \ .
\label{fstrip}
\eeq
This is analytic for all finite temperature, for both the ferromagnetic and
antiferromagnetic sign of the spin-spin coupling $J$.  The internal energy and
specific heat can be calculated in a straightforward manner from the free
energy (\ref{fstrip}); since the resultant expressions are somewhat cumbersome,
we do not list them here.  In contrast, for the range $1 < q < 2$, the
Potts antiferromagnet on the $L_x \to \infty$ limit of this open strip 
does have a phase transition at finite temperature (see eq. (\ref{tpstrip})),
but this has unphysical features, including a specific heat and partition 
function $Z$ that are negative for some range of temperature, and 
non-existence of a thermodynamic limit independent
of boundary conditions.  This will be discussed further below.

We next consider the limits of zero temperature for the antiferromagnetic and
ferromagnetic cases.  Let us define 
\beq
D_k(q) = \frac{P(C_k,q)}{q(q-1)} =
\sum_{s=0}^{k-2}(-1)^s {{k-1}\choose {s}} q^{k-2-s}
\label{dk}
\eeq
and $P(C_k,q)$ is the chromatic polynomial for the circuit
(cyclic) graph $C_k$ with $k$ vertices,
\beq
P(C_k,q) = (q-1)^k + (q-1)(-1)^k
\label{pck}
\eeq
so that $D_2=1$, $D_3=q-2$, $D_4=q^2-3q+3$, and so forth for other $D_k$'s.
In the $T=0$ Potts antiferromagnet limit $v=-1$,
$\lambda_{S,1}=D_3^2=(q-2)^2$ and $\lambda_{S,2}=0$, so that eq. 
(\ref{gammazcalc})
reduces to the generating function for the chromatic polynomial
for this open strip
\beq
\Gamma(S,q,-1;z) = \frac{q(q-1)(q-2)^2}{1-(q-2)^2z}
\label{gammacp}
\eeq
Equivalently, the chromatic polynomial is
\beq
P(S_m,q) = q(q-1)(q-2)^{2(m+1)} \ .
\label{pstrip}
\eeq

For the ferromagnetic case with general $q$, in the low-temperature limit
$v \to \infty$,
\beq
\lambda_{S,1} = v^4+4v^3+O(v^2) \ , \quad
\lambda_{S,2} = v^2+2(q-1)v+O(1) \quad {\rm as} \quad v \to \infty
\label{lamsvinf}
\eeq
so that $|\lambda_{S,1}|$ is never equal to $|\lambda_{S,2}|$ in this
limit, and hence ${\cal B}_u$ does not pass through the origin of the
$u$ plane for the $n \to \infty$ limit of the open square strip:
\beq
u=0 \not\in {\cal B}_u(\{S\}) .
\label{unotinbs}
\eeq
In contrast, as will be shown below, ${\cal B}_u$ does pass through
$u=0$ for this strip with cyclic or M\"obius boundary conditions.

\subsection{${\cal B}_q(\{S\})$ for fixed $a$}

\begin{figure}[hbtp]
\centering
\leavevmode
\epsfxsize=2.5in
\begin{center}
\leavevmode
\epsffile{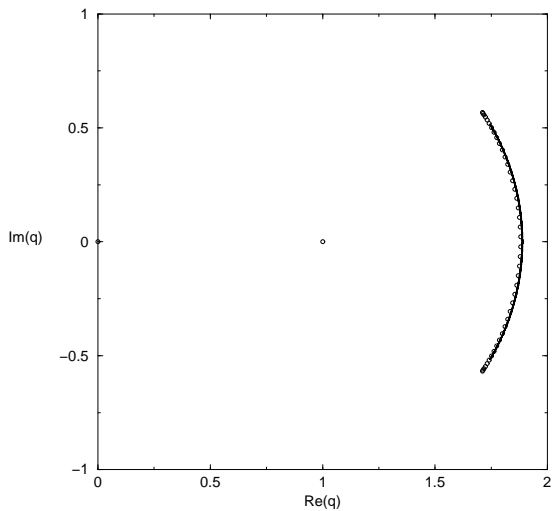}
\end{center}
\vspace{-10mm}
\caption{\footnotesize{Locus ${\cal B}_q$ for the $n \to \infty$ limit of the
$L_y=2$ triangular strip $\{S\}$ with free longitudinal boundary conditions, 
for $a=0.1$. Zeros of $Z(S_m,q,v=-0.9)$ for $m=20$ (i.e., $n=44$ vertices, so
that $Z$ is a polynomial of degree 44 in $q$) are shown for comparison.}}
\label{txy2a0p1}
\end{figure}

\begin{figure}[hbtp]
\centering
\leavevmode
\epsfxsize=2.5in
\begin{center}
\leavevmode
\epsffile{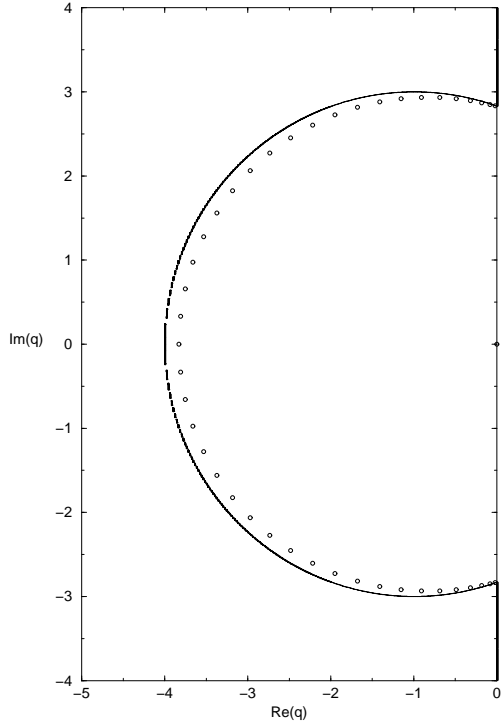}
\end{center}
\vspace{-5mm}
\caption{\footnotesize{Locus ${\cal B}_q$: same as Fig. \ref{txy2a0p1} for
$a=2$ (i.e., $v=1$).}}
\label{txy2a2}
\end{figure}

We start with the value $a=0$ corresponding to the Potts antiferromagnet at 
zero temperature.  In this case, $Z(S_m,q,v=-1)=P(S_m,q)$, where this chromatic
polynomial was given in eq. (\ref{pstrip}), 
and the continuous locus ${\cal B}_q=\emptyset$. For $a$
in the finite-temperature antiferromagnetic range $0 < a < 1$, ${\cal B}_q$
consists of a single self-conjugate arc crossing the positive real $q$ axis at 
\beq
q_c(\{S\}) = -v(3+v)=(1-a)(2+a)
\label{qcross}
\eeq
where the factor $(3v+v^2+q)$ multiplying the square root in eq. (\ref{lams})
vanishes, so that the two roots coincide. 
Parenthetically, we observe that this is same as the value of $q_c$ for the
Potts model on the $n \to \infty$ limit of the circuit graph, $C_n$
\cite{bcc,a}.  The arc endpoints occur at the branch points where the square
root is zero, viz., 
\beq
q_{S,endpt.} = (a-1)[a-2 \pm 2i\sqrt{a} \ ] \ . 
\label{qstripendpoint}
\eeq
As $a$ increases from 0 to 1, the crossing point (\ref{qcross}) decreases
monotonically from 2 to 0, and as $a$ reaches the infinite-temperature value 
1, ${\cal B}_q$ shrinks to a point at the origin.  In the ferromagnetic range
$a > 1$ the self-conjugate arc crosses the negative real axis, since 
$q_c(\{S\}) < 0$, and has support in the $Re(q) < 0$ half-plane.  In Figs. 
\ref{txy2a0p1} and \ref{txy2a2} we show ${\cal B}_q$ and associated zeros of
$Z$ in the $q$ plane for the antiferromagnetic value $a=0.1$ and the
ferromagnetic value $a=2$.  

\subsection{ ${\cal B}_u$ for fixed $q$}

\begin{figure}[hbtp]
\vspace{-20mm}
\centering
\leavevmode
\epsfxsize=2.5in
\begin{center}
\leavevmode
\epsffile{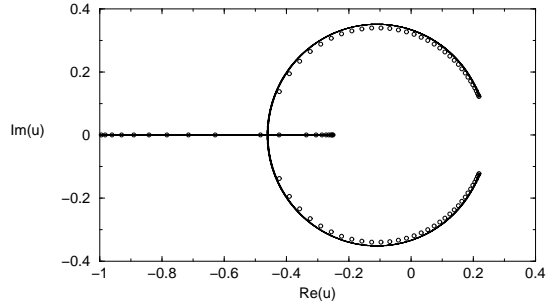}
\end{center}
\vspace{-20mm}
\caption{\footnotesize{Locus ${\cal B}_u$ for the $n \to \infty$ limit of the
$L_y=2$ triangular strip, with free longitudinal boundary conditions, $\{S\}$
with $q=10$. Zeros of $Z(S_m,q=10,v)$ in $u$ for $m=20$ are shown for 
comparison.}}
\label{txy2q10}
\end{figure}

\begin{figure}[hbtp]
\vspace{-20mm}
\centering
\leavevmode
\epsfxsize=2.5in
\begin{center}
\leavevmode
\epsffile{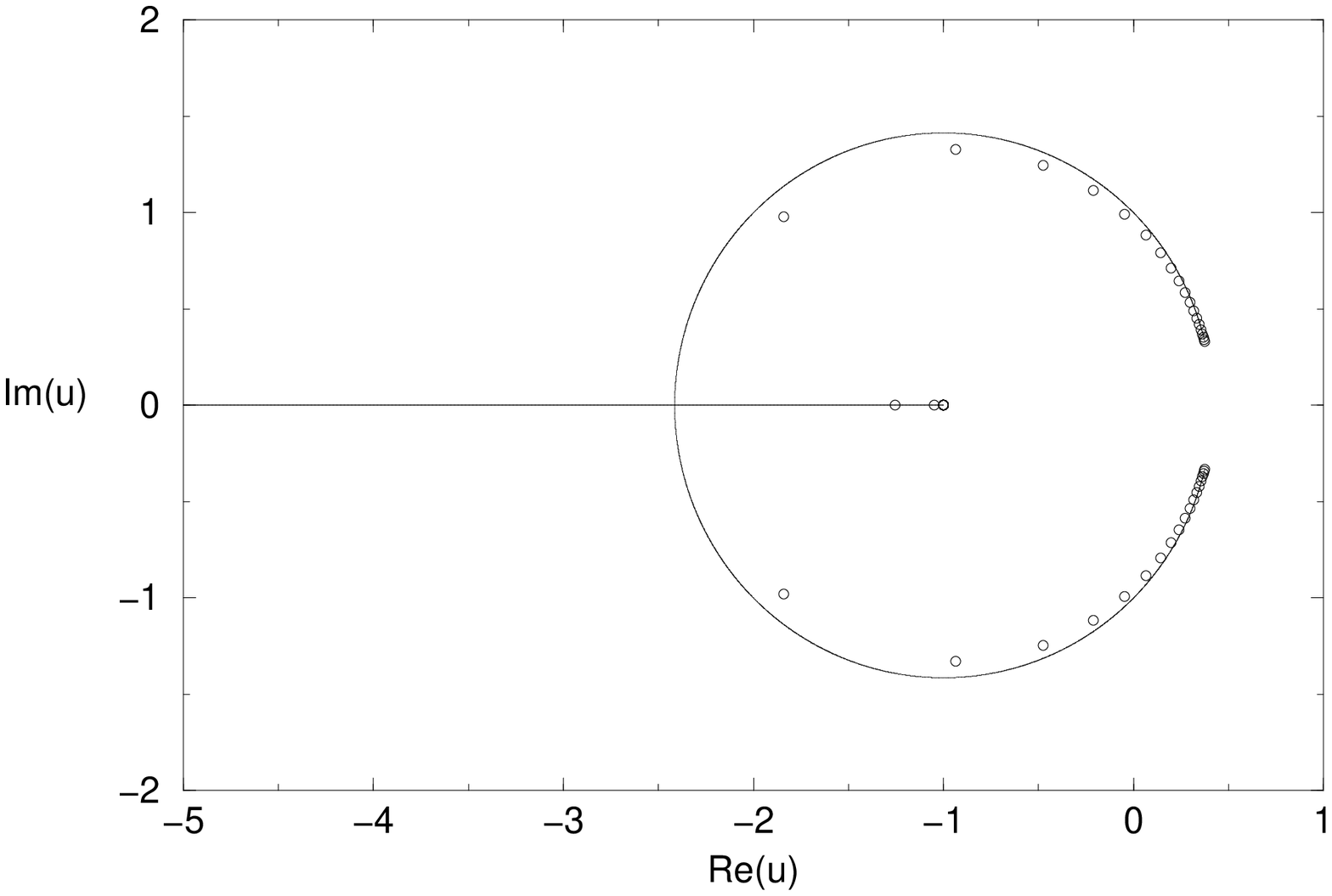}
\end{center}
\vspace{-20mm}
\caption{\footnotesize{Locus ${\cal B}_u$: same as in Fig. \ref{txy2q10} for
$q=2$.}}
\label{txy2q2}
\end{figure}

\begin{figure}[hbtp]
\vspace{-20mm}
\centering
\leavevmode
\epsfxsize=2.5in
\begin{center}
\leavevmode
\epsffile{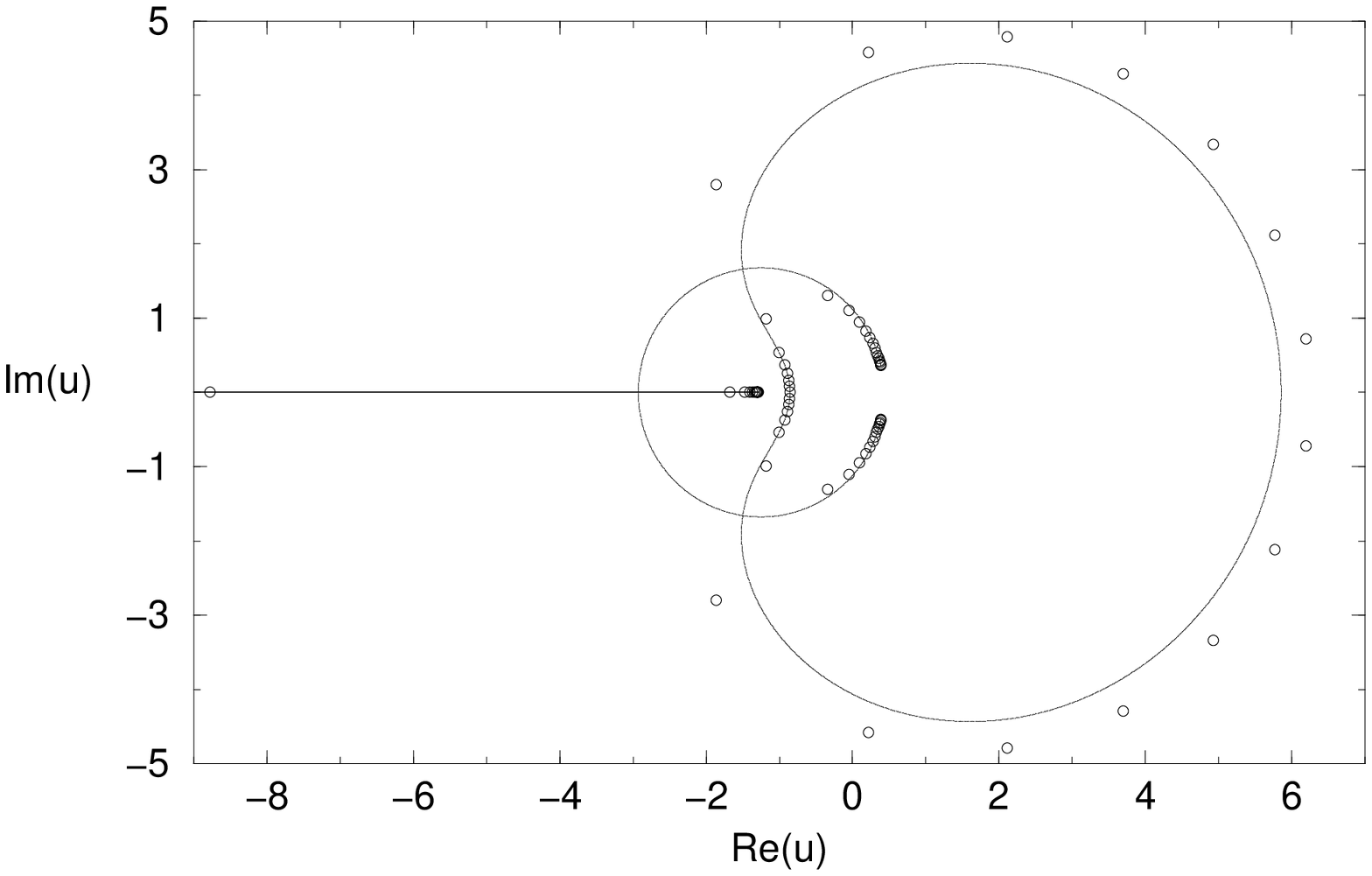}
\end{center}
\vspace{-20mm}
\caption{\footnotesize{Locus ${\cal B}_u$: same as in Fig. \ref{txy2q10} for
$q=1.8$.}}
\label{txy2q1p8}
\end{figure}

\begin{figure}[hbtp]
\vspace{-20mm}
\centering
\leavevmode
\epsfxsize2.5in
\begin{center}
\leavevmode
\epsffile{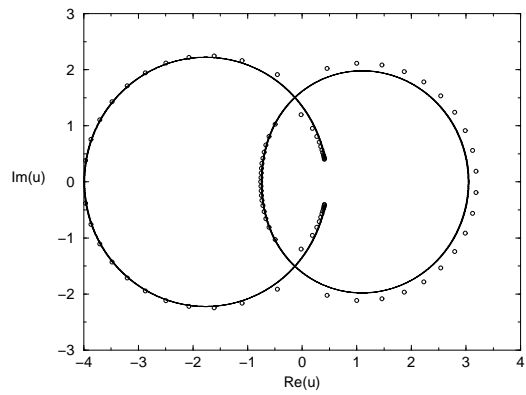}
\end{center}
\vspace{-20mm}
\caption{\footnotesize{Locus ${\cal B}_u$: same as in Fig. \ref{txy2q10} for
$q=(5/4)^2$.}}
\label{txy2qm}
\end{figure}

\begin{figure}[hbtp]
\vspace{-20mm}
\centering
\leavevmode
\epsfxsize=2.5in
\begin{center}
\leavevmode
\epsffile{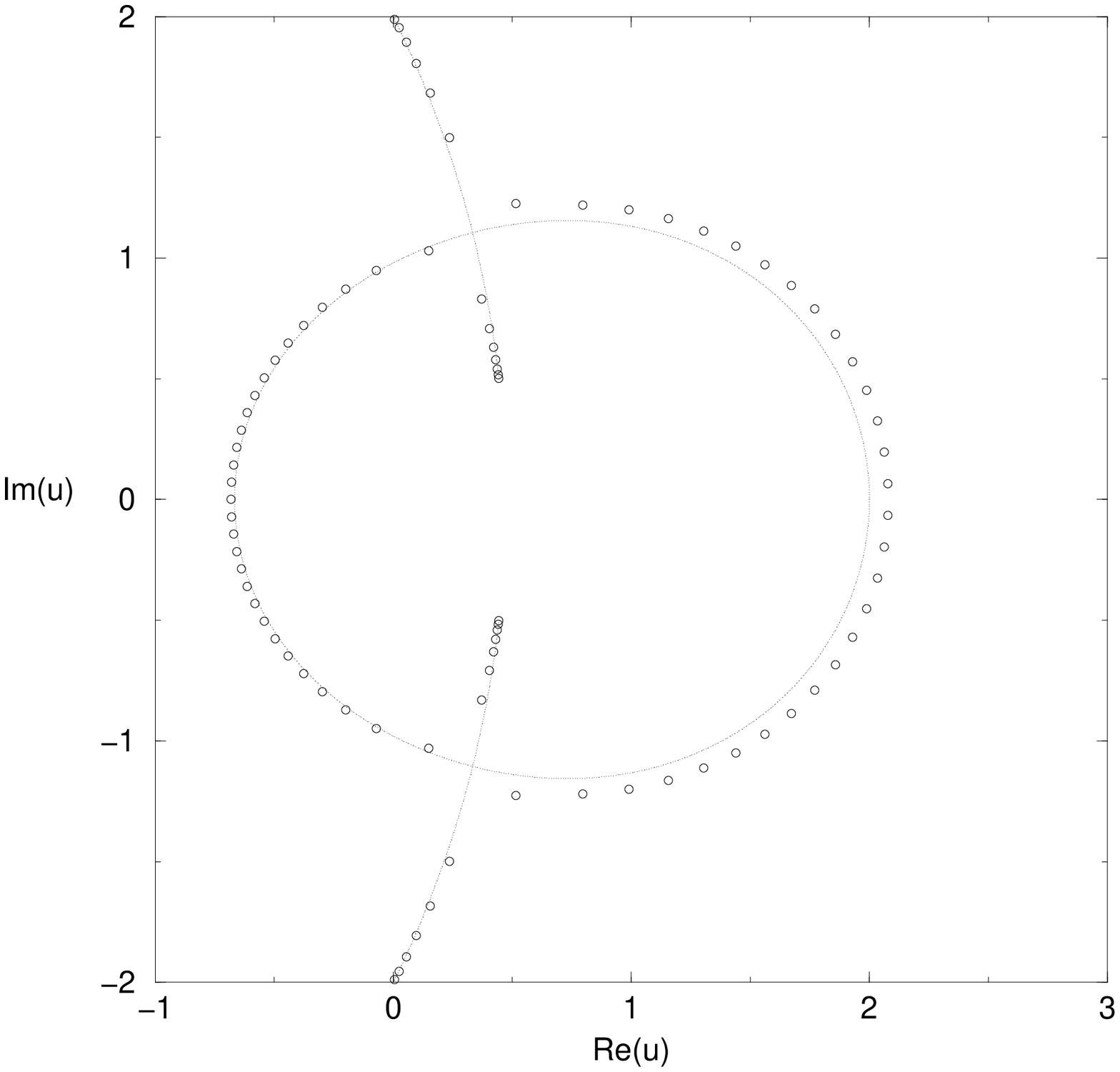}
\end{center}
\vspace{-10mm}
\caption{\footnotesize{Locus ${\cal B}_u$: same as in Fig. \ref{txy2q10} for
$q=1.25$.}}
\label{txy2q1p25}
\end{figure}

\begin{figure}[hbtp]
\vspace{-20mm}
\centering
\leavevmode
\epsfxsize=2.5in
\begin{center}
\leavevmode
\epsffile{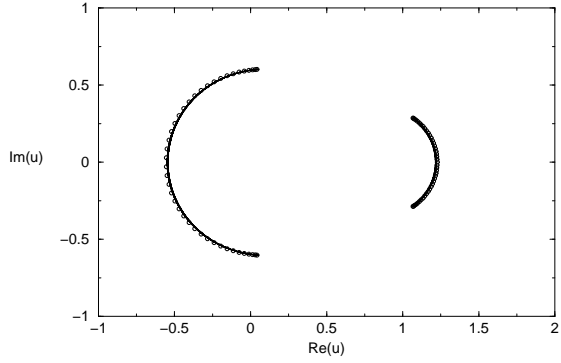}
\end{center}
\vspace{-20mm}
\caption{\footnotesize{Locus ${\cal B}_u$: same as in Fig. \ref{txy2q10} for
$q=0.5$.}}
\label{txy2q0p5}
\end{figure}

We show several plots of the locus ${\cal B}_u$ for various values of $q$ in 
Figs. \ref{txy2q10} - \ref{txy2q0p5}.  For values of $q$ where 
noncommutativity occurs, we consider ${\cal B}_{qn}$.  Given the algebraic 
structure of $\lambda_{S,j}$, $j=1,2$, the degeneracy of magnitudes 
$|\lambda_{S,1}|=|\lambda_{S,2}|$ and hence the locus ${\cal B}_u$ occurs 
where (i) $T_{S12}=0$, (ii) the prefactor of the square root vanishes: 
$3v+v^2+q=0$, (iii) $R_{S12}=0$, (iv) if $q$ is real, where $R_{S12} < 0$ so
that the square root is pure imaginary, and (v) elsewhere for complex $u$, 
where the degeneracy condition is satisfied.  All of these five possibilities
are realized here, although some, such as (i) can yield solutions already
subsumed by other conditions; for example, for the case $q=1.8$, shown in 
Fig. \ref{txy2q1p8}, $T_{S12}$ vanishes at two real points, 
$u \simeq -29.1, \ -1.35$, both of which are subsumed within the solution of 
condition (iv), which is a line segment $-69.486 < u < -1.291$, and so forth 
for various other solutions.  In cases where ${\cal B}_u$ does not
enclose regions, $\lambda_{S,1}$ is dominant everywhere in the $u$ plane, and
degenerate in magnitude with $\lambda_{S,2}$ on ${\cal B}_u$; the cases 
where ${\cal B}_u$ does enclose regions will be discussed individually. 

For large values of $q$, we find that ${\cal B}_u$ consists of the union of (i)
a circular arc centered at $u \sim -1/q$ that crosses the negative real $u$
axis, and (ii) a line segment on the negative $u$ axis.  For example, for
$q=10$, the arc has its endpoints at two of the branch point zeros of
$\sqrt{R_{S12}}$, at $u=(7 \pm \sqrt{15}i)/32  \simeq 0.21875 \pm 0.12103i$, 
and the line segment, which is the solution to the condition (iv) above, that
$R_{S12} < 0$ for real $q$, has left and right endpoints at $u_1 = -1$ and 
$u_2 = -1/4$.  As $q$ decreases, $u_1$ and $u_2$ move toward more
negative values, and as $q$ decreases to 2, $u_1 \to -\infty$, i.e. $a=0$,
while $u_2 \to -1$.  For $q=2$, ${\cal B}_u$ is the union of the above line
segment $-\infty < u < -1$ and the circular arc
\beq
u = -1 + \sqrt{2}e^{\pm i\theta} \ , \quad \theta_e \le \theta \le \pi
\label{ucircq2}
\eeq
where
\beq
\theta_e = arctan\Bigl ( \frac{\sqrt{7}}{11} \Bigr ) \simeq 13.5^\circ \ . 
\label{thetae}
\eeq
That is, the endpoints of the circular arc occur at
\beq
u_e = \frac{1}{8}(3 \pm \sqrt{7} \ i)
\label{ue}
\eeq

Proceeding, we observe that for $1 < q < 2$, the locus ${\cal B}_u$ 
separates the $u$ plane into different regions, while for $0 < q < 1$, 
as was true for  the range $q > 2$, ${\cal B}_u$ does not separate the $u$ 
plane into different regions.  In the range $q_m < q < 2$, where
\beq 
q=q_m = \Bigl ( \frac{5}{4} \Bigr )^2 = 1.5625 \ , 
\label{qm}
\eeq
${\cal B}_u$ crosses the real $u$ axis vertically at three different points and
separates the $q$ plane into three regions: (i) the paramagnetic phase, which
includes the infinite-temperature point $u=1$, where $\lambda_{S,1}$ is
dominant; (ii) the surrounding O phase that extends to the circle at infinity 
in the $u$ plane, i.e. to the point $a=0$, in which phase $\lambda_{S,2}$ is
dominant, and (iii) a third phase, of O type, located slightly to the left of 
$u=0$, in which $\lambda_{S,2}$ is dominant.  It is interesting that although
$\lambda_{S,2}$ is dominant in both of the O phases, there is still a boundary
that separates them completely; this is a result of the fact that on this
boundary, the other $\lambda$, namely $\lambda_{S,1}$, becomes degenerate in 
magnitude with $\lambda_{S,2}$.  Two of the points where ${\cal B}_u$ crosses
the real axis occur where the condition (ii) holds, i.e., where the 
prefactor $3v+v^2+q$ multiplying the square root in eq. (\ref{lams}), 
vanishes, at $v=(1/2)[-3 \pm \sqrt{9-4q} \ ]$, i.e., 
\beq
u = \frac{\Bigl [ 1 \pm \sqrt{9-4q} \ \Bigr ]}{2(2-q)} \ . 
\label{rell}
\eeq
For $q=2-\epsilon$ with $0 < \epsilon << 1$, these crossings are at 
$u=-1+O(\epsilon)$ and $u \sim 1/\epsilon$.  As $q$ decreases, the smaller
(larger) solution moves to the right (left).  Specifically, as $q$ decreases
from 2 to 0, the larger solution decreases monotonically from infinity to 1 
while the smaller one increases monotonically from $-1$ to $-1/2$.  For 
example, for $q=1.8$, the crossings in eq. (\ref{rell}) are at 
$u=(1/2)(5\pm 3\sqrt{5} \ ) \simeq -0.854, \ 5.854$. The existence of the 
right crossing on the positive real $u$ axis for $u > 1$ means that the 
free energy of the Potts antiferromagnet is nonanalytic at the temperature
\beq
T_{p,S} = \frac{J}{k_B \ln \Bigl [(1/2)(-1 + \sqrt{9-4q} \ ) \Bigr ]} 
\quad {\rm for} \quad 0 < q < 2
\label{tpstrip}
\eeq
(where both $J$ and the log are negative).  
However, just as found in \cite{a}, this nonanalyticity has associated 
unphysical features, including negative $Z$, negative specific heat in the 
low-temperature phase, and non-existence of a thermodynamic limit independent
of boundary conditions in the low-temperature phase.
As $q$ decreases from 2 to the value $q_m$ given above, 
the left and right endpoints of the line segment merge, and it shrinks
to a point at $u_1 = u_2 = -4$.  For this value $q=q_m$, the crossings in
eq. (\ref{rell}) are $u=(4/7)(2 \pm \sqrt{11} \ ) \simeq 3.038, \ -0.7524$. 
In Fig. \ref{txy2qm} we show ${\cal B}_u$ for $q=q_m$.  The arc endpoints 
occur at $u=(20 \pm 8\sqrt{6}i \ )/49 \simeq 0.4082 \pm 0.3999i$. 
For $q_m < q < 2$, there is a
multiple point on the negative real axis where a branch of ${\cal B}_u$ 
crosses this axis vertically and intersects with the line
segment.  There are also multiple points at complex-conjugate values
of $u$ where the circular arc intersects the closed curve.  As $q$
decreases from 2 to $q_m$, the circular arc in the left-hand complex
plane enlarges while the closed curve extending into the right-hand
plane shrinks and becomes convex. As $q$ decreases below $q_m$, the closed 
curve evident to the left in Fig. \ref{txy2qm} breaks apart, forming two 
complex-conjugate arcs, and in the interval $1 < q < q_m$, ${\cal B}_u$ 
no longer contains any line segment but instead consists of these
complex-conjugate arcs and the closed curve to the right; an illustrative 
example is given in Fig. \ref{txy2q1p25}, where the arc endpoints occur at 
$u=\pm 2i$ and $u = (1/9)(4 \pm 2\sqrt{5}i \ ) \simeq 0.444 \pm 0.497i$. 
As $q$ decreases toward 1, the arcs shrink to points at $u=e^{\pm i \pi/3}$.
At $q=1$, $({\cal B}_u)_{qn}$ is an oval (the solution to the degeneracy
equation $|u(1-u)|=1$) that (i) crosses the real axis at 
$u=(1/2)(1\pm \sqrt{5} \ )$, i.e. at approximately, 1.618 and $-0.618$, and
(ii) crosses the imaginary axis at 
$u=\pm[(1/2)(-1+\sqrt{5} \ )]^{1/2}i \simeq \pm 0.7862i$.  
For $0 < q < 1$, ${\cal B}$ consists only of two
disjoint arcs, as illustrated for the value $q=1/2$ in Fig. \ref{txy2q0p5}.
For $q=0$, the locus $({\cal B}_u)_{qn}$ is the circular arc 
\beq
u= \frac{1}{2}e^{i\theta} \ , \quad \frac{\pi}{2} \le \theta \le \frac{3\pi}{2}
\label{ucircq0}
\eeq
that crosses the real axis at $u=-1/2$ and has endpoints at $u = \pm i/2$. 
This is qualitatively similar to $({\cal B}_u)_{qn}$ for the $L_y=2$ open strip
of the square lattice at $q=0$, which was another circular arc crossing the
negative real axis at $u=-1/3$ with endpoints at $u=(-1 \pm 2\sqrt{2}i)/9$ 
\cite{a}. 

Note that at $q=0,1$, one encounters the noncommutativity (\ref{fnoncomm}); if
one sets $q$ to either of these values first and then takes $n \to
\infty$, the resultant ${\cal B}_u=\emptyset$.

Our findings for this $L_y=2$ open strip of the triangular lattice may be
contrasted with those for the $L_y=2$ open strip of the square lattice in 
\cite{a}.  In the latter case, ${\cal B}_u$ consisted of arcs that never
enclosed any regions for real $q \ge 0$.  

\newpage

\section{Cyclic and M\"obius Strips of the Triangular Lattice} 

\subsection{Results for $Z$}

By either using an iterative application of the deletion-contraction theorem
for Tutte polynomials and converting the result to $Z$, or by using a
transfer matrix method (in which one starts with a $q^2 \times q^2$
transfer matrix and generalizes to arbitrary $q$), one can calculate the
partition function for the cyclic and M\"obius ladder graphs of arbitrary
length, $Z(G,q,v)$, $G=L_m, \ ML_m$.  We have used both methods as checks on
the calculation.  Our results have the general form (\ref{zgsum}) with 
$N_\lambda=6$: 
\beq
Z(G_m,q,v) = \sum_{j=1}^6 c_{G,j}(\lambda_{G,j}(q,v))^m \ , G_m = L_m, ML_m
\label{zgsuml}
\eeq
where 
\beq
\lambda_{L,j}=\lambda_{ML,j} \ , j=1,...,6
\label{lamlml}
\eeq
We find 
\beq
\lambda_{L,1}=v^2
\label{lam1}
\eeq
and 
\beq
\lambda_{L,5}=\lambda_{S,1} \ , \quad 
\lambda_{L,6}=\lambda_{S,2} \ . 
\label{lam56}
\eeq
The $\lambda_j$ for $j=2,3,4$, are the solutions of the cubic equation 
\beqs
& & \eta^3-v(v^3+8v+4v^2+2q)\eta^2 + 
v^2(2v^3q+6v^3+8qv+q^2+8v^2+2v^4+6qv^2)\eta \cr\cr
& - & v^4(v+1)^2(v+q)^2=0 \ . 
\label{cubicz}
\eeqs
For the cyclic strip the coefficients in eq. (\ref{zgsuml}) are 
\beq
c_{L,1} = q^2-3q+1
\label{c1}
\eeq
\beq
c_{L,j}=q-1 \quad {\rm for} \quad j=2,3,4
\label{c234}
\eeq
\beq
c_{L,j}=1 \quad {\rm for} \quad j=5,6
\label{c56}
\eeq
For the M\"obius strip, we have 
\beq
c_{ML,1}=-1
\label{cml1}
\eeq
and 
\beq
c_{ML,j}=1  \quad {\rm for} \quad j=5,6 \ . 
\label{cml56}
\eeq
The coefficients $c_{ML,j}$, $j=2,3,4$ are more complicated and are given 
by the generating function (see the appendix) by the formula 
given in \cite{a}.  

Our exact calculations (cf. eq. (\ref{lamlml})) yield the following general 
result
\beq
{\cal B}(\{L\}) = {\cal B}(\{ML\}) \ .
\label{bcycmob}
\eeq 
This is the same result that one of us found for the analogous strip of the
square lattice \cite{bcc,a} and is in accord with the conclusion that the
singular locus is the same for an infinite-length finite-width strip graph for
given transverse boundary conditions, independent of the longitudinal boundary
condition.  Owing to the equality (\ref{bcycmob}), we shall henceforth, for
brevity of notation, refer to both ${\cal B}(\{L\})$ and ${\cal B}(\{ML\})$ as
${\cal B}(\{L\})$ and similarly for specific points on ${\cal B}$, such as
$q_c(\{L\})=q_c(\{ML\})$, etc.

Our main interest here is in large $m$ and the $m \to \infty$ limit.  However,
for completeness, we make the following remark.  If $m \ge 3$, then $L_m$ is a
(proper) graph, but the $m=1$ and $m=2$ cases requires special consideration;
in these cases, $L_m$ degenerates and is not a proper
graph\footnote{\footnotesize{A proper graph has no multiple edges or loops,
where a loop is an edge that connects a vertex to itself.  A multigraph may
contain multiple edges, but no loops, while a pseudograph may contain both
multiple edges and loops \cite{bbook,boll}.}}.  $L_2$ is the multigraph
obtained from the complete graph $K_4$ \footnote{\footnotesize{The complete
graph $K_p$ is the graph with $p$ vertices each of which is adjacent to all of
the other vertices.}} by doubling two non-adjacent edges (i.e., edges that do
not connect to any common vertex). $L_1$ is
the pseudograph obtained by connecting two vertices with a double edge and
adding a loop to each vertex.  Our calculation of $Z(L_m,q,v)$ and the
corresponding Tutte polynomial $T(L_m,x,y)$ apply not just for the uniform
cases $m \ge 3$ but also for the special cases $m=1,2$ if 
for $m=2$ one includes the multiple edges and for $m=1$ the
multiple edges and loops in the evaluation of (\ref{zfun}), (\ref{ham}), and
(\ref{cluster}).  Note that in the $T=0$ case for the antiferromagnet, the
resulting partition function, or equivalently, the chromatic polynomial, is not
sensitive to multiple edges, i.e. is the same for a graph in which two vertices
are connected by one edge or multiple edges; however, the general partition
function (Tutte polynomial) is sensitive to multiple edges.  The chromatic
polynomial is sensitive to loops and vanishes identically when a pseudograph
has any loops.

\subsection{Special values and expansions of $\lambda$'s}

We discuss some special cases. First, for the zero-temperature Potts
antiferromagnet, i.e. the case $a=0$ ($v=-1$), the partition functions
$Z(L_m,q,v)$ and $Z(ML_m,q,v)$ reduce, in accordance with the general result
(\ref{zp}), to the respective chromatic polynomials $P(L_m,q)$ and $P(ML_m,q)$
\cite{wcy} with $\lambda_j$'s comprised of the four terms 1, $(q-2)^2$, and
$(1/2)[5-2q \pm \sqrt{9-4q}]$.  The two remaining $\lambda_{L,j}$'s vanish in
this limit.  (Since only four $\lambda_{L,j}$'s occur in the chromatic
polynomial, a different numbering convention was used in \cite{wcy} than here,
where, in general, six occur.) For the infinite-temperature value $a=1$, we
have $\lambda_{L,j}=0$ for $j=1,2,3,4,6$, while $\lambda_{L,5}=q^2$, so that
$Z(G,q,a=1)= q^{2m} = q^n$ for $G=L_m, \ ML_m$, in accord with the general
result (\ref{za1}).  For the real interval $q \ge 3$ and the region $R_1$ to
which one can analytically continue from this interval (see Fig. 
\ref{tpxy2a0} below), $W=q-2$.  Hence, $W=1$ at $q=3$.  A technical remark is
the following: one can, and it is convenient to, take the $n \to \infty$ limit
with $m=0$ mod 3, so that, by eqs. (\ref{chitly2}) and (\ref{pq3}), $P=3!$, so
that the limit for $W$ exists at $a=0$ as well as at $a \ne 0$.  In contrast,
if one took $n \to \infty$ using all positive integer values of $m$, then, 
strictly speaking, the limit for $W$ in eq. (\ref{w}) would not exist for 
$a=0$, since $P$ would have the nonconvergent values $6,0,0,6,,..$ for 
$m=3k,3k+1,3k+2,3k+3,..$  For the M\"obius longitudinal boundary condition, no
such convenient choice is possible, since $\chi=4$ for all $m$; here there is a
noncommutativity: if one starts with $q$ slightly larger than 3, takes $n \to
\infty$ to calculate $W$, and then lets $q \searrow 3$, one gets $W(q=3)=1$, 
but if one sets $q=3$ first and then lets $n \to \infty$, one gets $W(q=3)=0$.

At $q=0$ (with appropriate choices of branch cuts) we find that 
\beq
\lambda_{L,1}=\lambda_{L,3}=v^2
\label{lam1lam3q0}
\eeq
\beq
\lambda_{L,2}=\lambda_{L,5}=
\frac{v^2}{2}\Bigl [ v^2+4v+7 + (v+3)\sqrt{v^2+2v+5} \ \Bigr ] 
\label{lam2lam5q0}
\eeq
and
\beq
\lambda_{L,4}=\lambda_{L,6}=
\frac{v^2}{2}\Bigl [ v^2+4v+7 - (v+3)\sqrt{v^2+2v+5} \ \Bigr ] \ . 
\label{lam3lam6q0}
\eeq
Since there are dominant terms that are degenerate, namely 
$\lambda_{L,2}=\lambda_{L,5}$, it follows that 
\beq
q=0 \quad {\rm is \ on} \quad {\cal B}_q(\{L\}) \quad \forall \ a \ .
\label{q0onb}
\eeq
This was also true of the circuit graph and cyclic and M\"obius square strips
with $L_y=2$ for which the general Potts model partition function (Tutte
polynomial) was calculated in \cite{a}.  For $q=0$, the coefficients
$c_j=1$ for $j=1,5,6$ and $c_j=-1$ for $j=2,3,4$ so that the equal terms
cancel each other pairwise, yielding $Z(L_m,q=0,v)=0$, in accordance with 
the general result (\ref{zq0}).  The noncommutativity (\ref{fnoncomm}) occurs 
here: $\exp(f_{nq})=0$, while $|\exp(f_{qn})|=|\lambda_{L,5}|^{1/2}$. 

At $q=1,2$ we again encounter noncommutativity in the calculation of
the free energy.  For $q=1$, $f_{nq}=2\ln a = 2K$, while $f_{qn}$ depends on
which phase one is in for a given value of $a$.  For $0 < a < a_c(q=1)$, where,
from eq. (\ref{acl}), $a_c(q=1)=(1/2)(-1+\sqrt{5}) \simeq 0.6180$, 
$f_{qn}=(1/2)\ln \lambda_{L,c}$, where $\lambda_{L,c}$ is the cube root that is
dominant in this phase (corresponding to $(1/2)[5-2q+\sqrt{9-4q}]$ for $a \to
0$), while for $a > a_c(q=1)$, $f_{qn}=(1/2)\ln \lambda_{L,5}$. 

Similarly, for $q=2$, again with an appropriate choice of
branch cuts, 
\beq
\lambda_{L,1}=\lambda_{L,3}=v^2
\label{lam1lam3q2}
\eeq
\beq
\lambda_{L,(2,4)}=\frac{v(v+1)}{2}\biggl [ v^2+3v+4 \pm \Bigl [
v(v+1)(v^2+5v+8) \Bigr ]^{1/2} \biggr ]
\label{lam24q2}
\eeq
and 
\beq
\lambda_{L,(5,6)}=\frac{(v+1)(v+2)}{2}\biggl [(v^2+v+2) \pm \Bigl [
(v+1)(v+2)(v^2-v+2) \Bigr ]^{1/2}  \biggr ] \ .
\label{lam56q2}
\eeq
For this value, $q=2$, the coefficients are 
$c_{L,1}=-1$, while $c_{L,j}=1$, $2 \le j \le 6$; hence,
the $(\lambda_{L,1})^m$ and $(\lambda_{L,3})^m$ terms cancel each other and
make no contribution to $Z$, which reduces to
\beq
Z(L_m,q=2,v)=\sum_{j=2,4,5,6} (\lambda_{L,j})^m
\label{zq2}    
\eeq
Hence also, $f_{qn} \ne f_{nq}$ at $q=2$. 

We observe that the $\lambda_j$'s have a more symmetric form for $q=2$ when
expressed in terms of the variable $u$:
\beq
\lambda_{L,1,u}=\lambda_{L,3,u}=u^2(1-u)^2
\label{lam13q2u}
\eeq
\beq
\lambda_{L,(2,4),u}=\frac{(1-u)}{2}\biggl [ 1+u+2u^2 \pm \Bigl [
(1-u)(1+3u+4u^2) \Bigr ]^{1/2} \biggr ]
\label{lam24q2u}
\eeq
and 
\beq
\lambda_{L,(5,6),u}=\frac{(1+u)}{2}\biggl [ 1-u+2u^2 \pm \Bigl [
(1+u)(1-3u+4u^2) \Bigr ]^{1/2} \biggr ] \ . 
\label{lam56q2u}
\eeq
One sees that each member of the pair $\lambda_{L,(2,4),u}$ is equal to 
the respective member of the pair $\lambda_{L,(5,6),u}$ with the replacement 
$u \to -u$.  It follows that $|\lambda_{L,2,u}|=|\lambda_{L,5,u}|$ and 
$|\lambda_{L,4,u}|=|\lambda_{L,6,u}|$ for $u$ pure imaginary.

In general, for $q \in {\mathbb Z}_+$, the partition function $Z(L_m,q,v)$ for
the cyclic width $L_y=2$ strip of the triangular lattice is identical to the
partition function for the 1D $q$-state Potts model with nearest-neighbor and
next-nearest-neighbor spin-spin couplings that are equal in magnitude.  This
equality can be seen easily by redrawing the strip of the triangular lattice as
a line with additional couplings between next-nearest-neighbor vertices on this
line.  In Ref. \cite{1dnnn} Tsai and one of us calculated $Z$ for the latter
model (with, in general, unequal nearest and next-nearest-neighbor spin-spin
couplings).  Hence, in particular, the $q=2$ and $q=3$ of eq. (\ref{zgsuml}) 
coincide with the results in (section IX of) \cite{1dnnn}.  

In order to study the zero-temperature critical point in the ferromagnetic
case and also the properties of the complex-temperature phase diagram, we
calculate the $\lambda_{G,j,u}$'s corresponding to the $\lambda_{G,j}$'s,
using eq. (\ref{lamu}).  In the vicinity of the point $u=0$ we have
\beq
\lambda_{L,1,u}=u^2(1-u)^2  
\label{lam1rtaylor}
\eeq
and the Taylor series expansions
\beq
\lambda_{L,2,u}=1 - 2u^3 + 2(q-1)u^4 + O(u^5) 
\label{lam2rtaylor}
\eeq
\beq
\lambda_{L,3,u}=u^2 + O(u^3) 
\label{lam3rtaylor}
\eeq
\beq
\lambda_{L,4,u}=u^2 + O(u^3) 
\label{lam4rtaylor}
\eeq
\beq
\lambda_{L,5,u}=1+2(q-1)u^3\Bigl [ 1 + u + O(u^2) \Bigr ]
\label{lam5rtaylor}
\eeq
\beq
\lambda_{L,6,u}=u^2 + 2(q-2)u^3 + O(u^4) \ . 
\label{lam6rtaylor}
\eeq
Hence, at $u=0$, $\lambda_{L,2,u}$ and $\lambda_{L,5,u}$ are dominant and
$|\lambda_{L,2,u}|=|\lambda_{L,5,u}|$, so that the point $u=0$ is on
${\cal B}_u$ for any $q \ne 0,1$, where the noncommutativity (\ref{fnoncomm})
occurs.  
To determine the angles at which the branches of ${\cal B}_u$ cross
each other at $u=0$, we write $u$ in polar coordinates as $u=re^{i\theta}$,
expand the degeneracy equation $|\lambda_{L,2,u}|=|\lambda_{L,5,u}|$, for small
$r$, and obtain $qr^3\cos(3\theta)=0$, which implies that (for $q \ne 0,1$) in
the limit as $r=|u| \to 0$,
\beq
\theta = \frac{(2j+1)\pi}{6} \ , \quad j=0,1,...,5
\label{thetau}
\eeq
or equivalently, $\theta=\pm \pi/6$, $\pm \pi/2$, and $\theta=\pm 5\pi/6$.
Hence there are six curves forming three branches of ${\cal B}_u$ 
intersecting at $u=0$ and successive branches cross at an angle of $\pi/3$. 
The point $u=0$ is thus a multiple point on
the algebraic curve ${\cal B}_u$, in the technical terminology of algebraic
geometry (i.e., a point where several branches of an algebraic curve cross
\cite{alg}).  In the vicinity of the origin, $u=0$, these branches define six
complex-temperature phases: the paramagnetic (PM) phase for 
$-\pi/6 \le \theta \le \pi/6$, together with the phases $O_j$ for $1 \le j \le
5$, with $O_j$ occupying the sector $(2j-1)\pi/6 \le \theta \le (2j+1)\pi/6$.
Note that $O_3=O_3^*$, $O_4=O_2^*$, and $O_5=O_1^*$.  
For the case of interest here, namely, $q > 0$, $\lambda_{L,5,u}$ is dominant 
in the PM phase and in the $O_2$ and $O_2^*$ phases, while $\lambda_{L,2,u}$ is
dominant in the $O_1$, $O_1^*$, and $O_3$ phases. 

One of our interesting findings is that for $q=2$ and for $q=3$ the Potts
antiferromagnet on the infinite-length, width $L_y=2$ strip of the triangular
lattice has a zero-temperature critical point.  (As must be the case for this
to be physical, this is independent of the longitudinal boundary conditions.)
In the $q=2$ Ising case, this involves frustration, and the 2D Ising
antiferromagnet on the triangular lattice also has a $T=0$ critical point
\cite{istri}.  In contrast, for the $q=3$ case, the $T=0$ critical point does
not involve any frustration, and the $q$ value at which this occurs for our
$L_y=2$ strip is different than the value, $q=4$, where the Potts
antiferromagnet has a $T=0$ critical point on the full 2D triangular lattice.
In order to study the $T=0$ critical point for the $L_y=2$ strip for these two
values of $q$, it is useful to calculate expansions of the $\lambda_j$'s; only 
$\lambda_{L,5}$ and $\lambda_{L,2}$ are necessary for physical thermodynamic 
properties, while the full set of $\lambda_{L,j}$, $j=1,2,..,6$ is necessary 
for the study of the singular locus ${\cal B}$.

For $q=2$, besides the exact expressions 
$\lambda_{L,1}=\lambda_{L,3}=(1-a)^2$ from eq. (\ref{lam1lam3q2}), we have the
expansions 
\beq
\lambda_{L,(2,4)} = -a + \frac{a^2}{2}(1+a^2) \pm i\sqrt{a}\Bigl [ -a + 
\frac{9}{8}a^2 +O(a^3) \Bigr ]
\label{lam24q2taylor}
\eeq
and
\beq
\lambda_{L,(5,6)} = a + \frac{a^2}{2}(1+a^2) \pm \sqrt{a}\Bigl [a + 
\frac{9}{8}a^2 +O(a^3) \Bigr ] \ . 
\label{lam56q2taylor}
\eeq
Note that (i) these are not Taylor series expansions in $a$, but rather in the
variable $\sqrt{a}$, and (ii) each member of the pair $\lambda_{L,(2,4)}$ is
equal to the respective member of the pair $\lambda_{L,(5,6)}$ with the
replacement $a \to -a$.   As shown above, for $f_{nq}$ and 
${\cal B}_{nq}$, where one sets $q=2$ first and then takes $n \to \infty$,
$\lambda_{L,j}$, $j=1,3$, make no contribution, and  ${\cal B}_{nq}$ is
determined by the degeneracy in magnitude of $\lambda_{L,j}$, $j=2,4,5,6$.
 From the expansions (\ref{lam24q2taylor}) and (\ref{lam56q2taylor}), it
follows that in the neighborhood of the point $a=0$, $({\cal B}_a)_{nq}$ is
determined by the equation $|\lambda_{L,2}|=|\lambda_{L,5}|$ and 
is a vertical line segment along the imaginary $a$ axis. This will be 
discussed further below (see Fig. \ref{tpxy2aq2}). 

For $q=3$, besides the exact expression $\lambda_{L,1}=(1-a)^2$,
we calculate the expansions 
\beq
\lambda_{L,(2,3)} = e^{\pm 2\pi i/3} -2e^{\pm \pi i/3}a + \Bigl ( -1 \pm 
\frac{2i\sqrt{3}}{3} \ \Bigr ) a^2 + O(a^3) 
\label{lam23q3taylor}
\eeq
(where the upper (lower) sign applies for $j=2$ ($j=3$)); 
\beq
\lambda_{L,4} = 4a^2 - 4a^3 +O(a^4)
\label{lam4q3taylor}
\eeq
\beq
\lambda_{L,5} = 1 + 6a - 3a^2 + O(a^3) 
\label{lam5q3taylor}
\eeq
and
\beq
\lambda_{L,6} = 4a^2 - 28a^3 +O(a^4) \ . 
\label{lam6q3taylor}
\eeq
There are thus four dominant $\lambda_{L,j}$'s at $a=0$, viz., those with
$j=1,2,3,5$, which satisfy $|\lambda_{L,j}|=1$.  Writing $a=\rho e^{i\phi}$ and
expanding the dominant $\lambda_{L,j}$'s in the neighborhood of $a=0$, we
obtain
\beq
|\lambda_{L,1}|^2 = 1-4\rho\cos\phi +O(\rho^2)
\label{lam1polar}
\eeq
\beq
|\lambda_{L,(2,3)}|^2 = 1-4\rho\cos(\phi \mp \pi/3) +O(\rho^2)
\label{lam23polar}
\eeq
and
\beq
|\lambda_{L,5}|^2 = 1+12\rho\cos\phi +O(\rho^2) \ . 
\label{lam5polar}
\eeq
Equating dominant $|\lambda_{L,j}|$'s, it follows that in the neighborhood of
$a=0$, ${\cal B}_a$ consists of four curves, forming complex-conjugate pairs,
passing through $a=0$ at the angles
\beq
\phi = \pm arctan \Big ( \frac{7\sqrt{3}}{3} \Bigr ) \simeq 76.10^\circ \ . 
\label{phi1}
\eeq
and 
\beq
\phi = \pm \frac{5\pi}{6} = \pm 150^\circ
\label{phi2}
\eeq
In Fig. \ref{tpxy2aq3} below we show ${\cal B}_a$ for this case.

\subsection{${\cal B}_q(\{L\})$ for fixed $a$}

\subsubsection{Antiferromagnetic case $0 \le a \le 1$} 

\begin{figure}[hbtp]
\centering
\leavevmode
\epsfxsize=2.5in
\begin{center}
\leavevmode
\epsffile{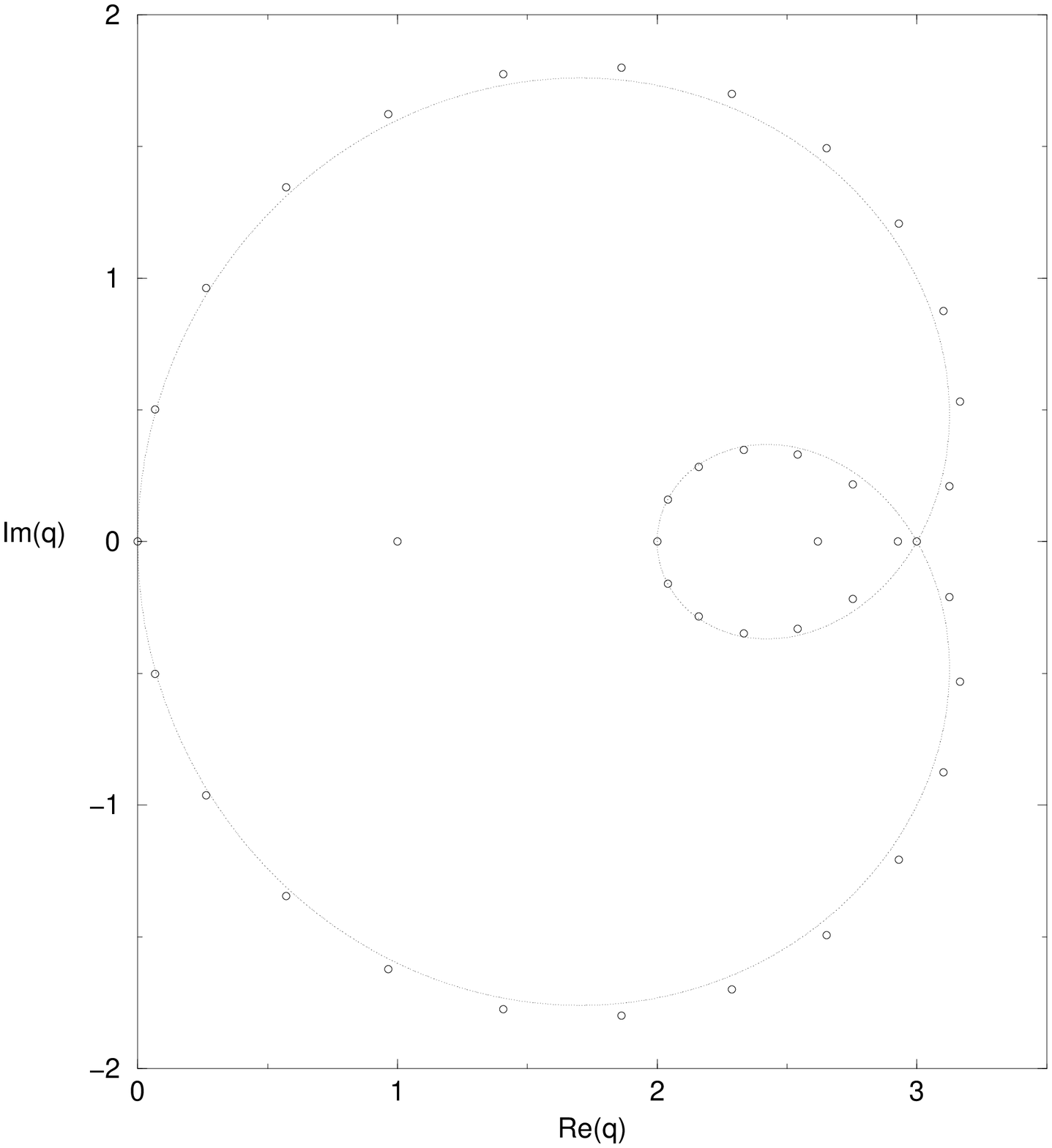}
\end{center}
\vspace{-5mm}
\caption{\footnotesize{Locus ${\cal B}_q(\{L\})$ for the $n \to \infty$ limit
of the $L_y=2$ cyclic or M\"obius triangular strip for $a=0$, i.e., the $T=0$
limit of the Potts antiferromagnet. Chromatic zeros for $m=20$ (so that $Z$ is
a polynomial of degree $n=40$ in $q$) are shown for comparison.}}
\label{tpxy2a0}
\end{figure}

We begin with the case $a=0$, i.e. the $T=0$ limit of the Potts
antiferromagnet.  The locus ${\cal B}_q$ for this case was stated in
\cite{wcyl} and given as Fig. 3 in \cite{wcy}.  
This locus ${\cal B}_q$, shown as Fig. \ref{tpxy2a0} here, 
separates the complex $q$
plane into three different regions: (i) $R_1$, including the segments $q > 3$
and $q < 0$ on the real axis, where $W=(q-2)$; (ii) $R_2$, including the real 
interval $2 < q < 3$, in which $|W|=1$; and (iii) $R_3$, including the
real segment $0 < q < 2$, in which $|W|=[(1/2)(5-2q+\sqrt{9-4q} \ ) ]^{1/2}$.
At $q=q_c$ there are actually four terms that are degenerate in magnitude, 
\beq
\lambda_{L,1}=\lambda_{L,5}=|\lambda_{L,2}|=|\lambda_{L,3}|=1 \quad {\rm at}
\quad q=3 \quad {\rm for} \quad a=0
\label{degeneq_q3}
\eeq
corresponding to the property that this is a multiple point on ${\cal B}_q$ (in
the terminology of algebraic geometry) where four curves intersect. 
In contrast, the other two points at which 
${\cal B}_q$ crosses the real axis, at $q=0$ and $q=2$, involve only the
minimal number (two) of degenerate magnitudes and are hence not multiple
points.  Evidently, $q_c(\{L\})=3$ for this $a=0$ case.

\begin{figure}[hbtp]
\centering
\leavevmode
\epsfxsize=2.5in
\begin{center}
\leavevmode
\epsffile{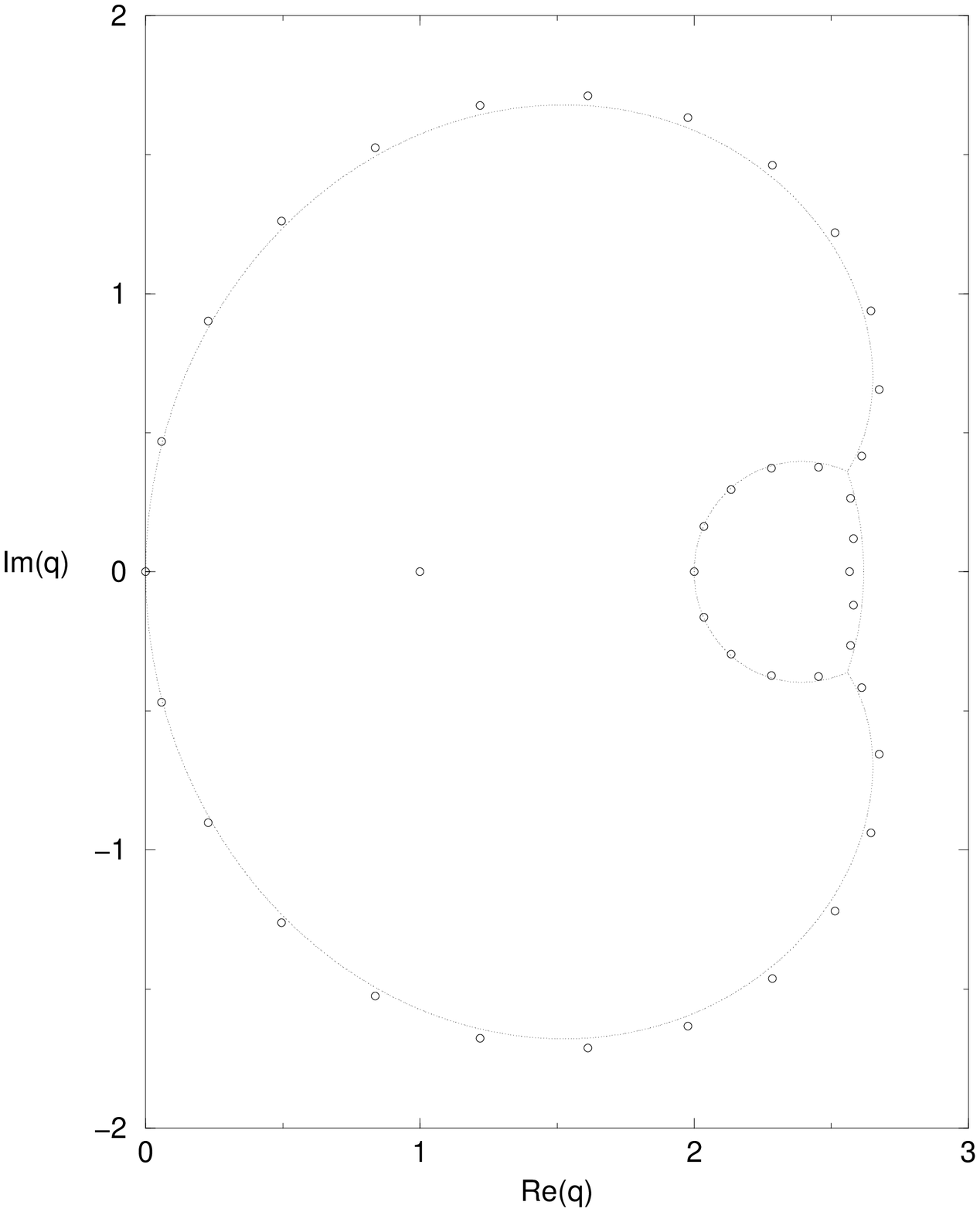}
\end{center}
\vspace{-5mm}
\caption{\footnotesize{Locus ${\cal B}_q(\{L\})$: same as Fig. \ref{tpxy2a0}
for $a=0.1$ (finite-temperature Potts antiferromagnet).}}
\label{tpxy2a0p1}
\end{figure}

\begin{figure}[hbtp]
\centering
\leavevmode
\epsfxsize=2.5in
\begin{center}
\leavevmode
\epsffile{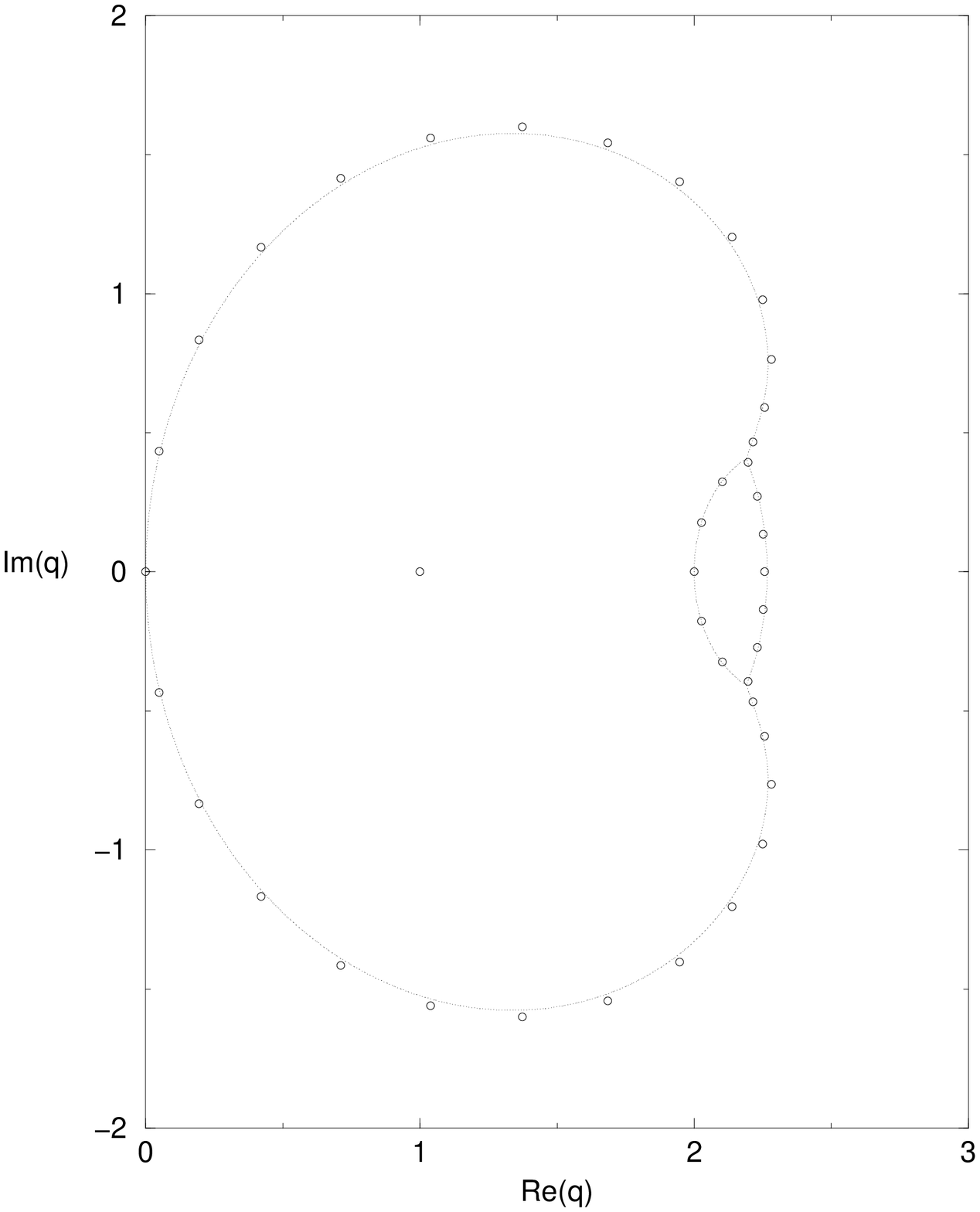}
\end{center}
\vspace{-5mm}
\caption{\footnotesize{Locus ${\cal B}_q(\{L\})$: same as Fig. \ref{tpxy2a0}
for $a=0.2$.}}
\label{tpxy2a0p2}
\end{figure}

\begin{figure}[hbtp]
\centering
\leavevmode
\epsfxsize=2.5in
\begin{center}
\leavevmode
\epsffile{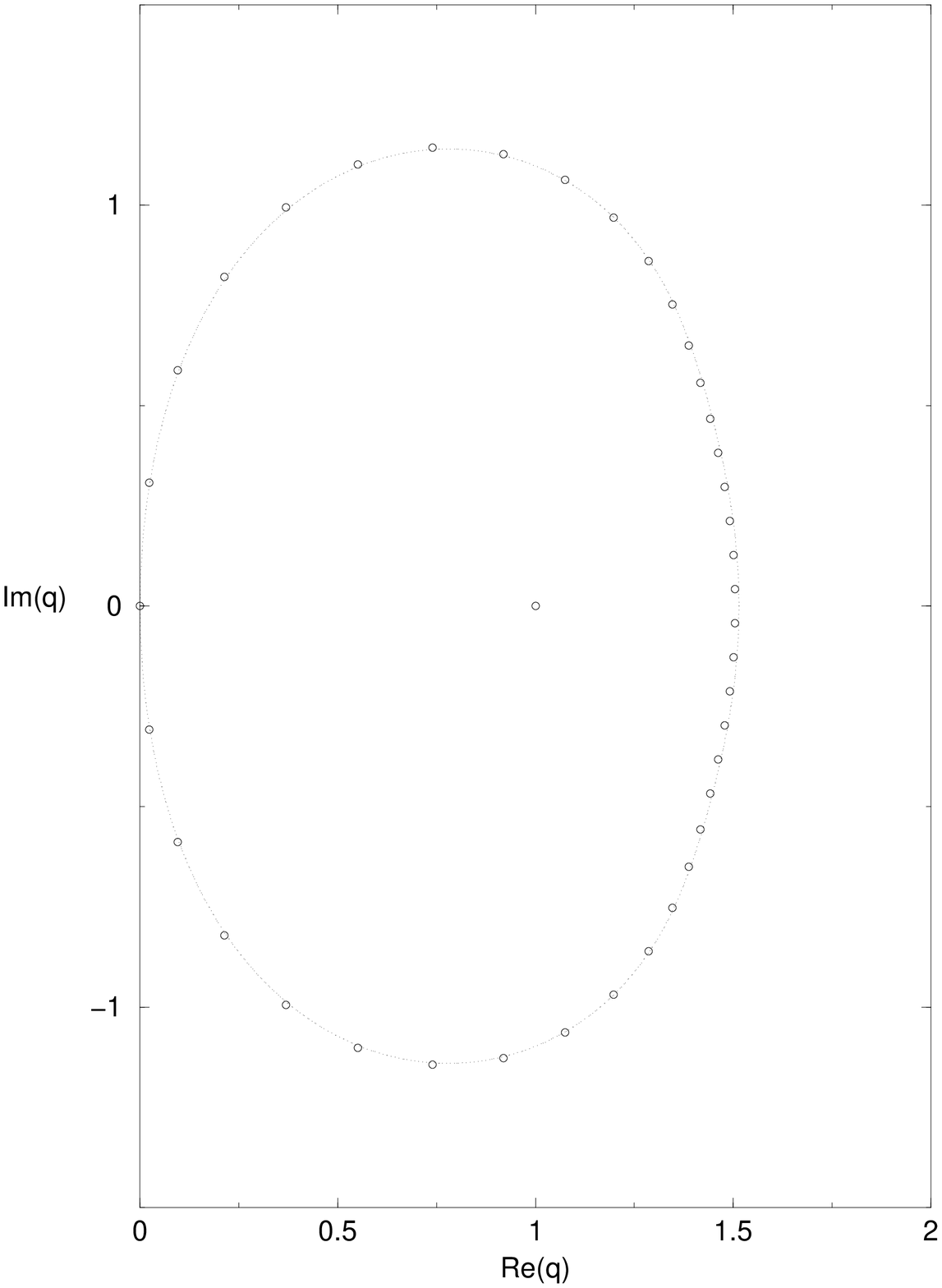}
\end{center}
\vspace{-5mm}
\caption{\footnotesize{Locus ${\cal B}_q(\{L\})$: same as Fig. \ref{tpxy2a0}
for $a=0.5$.}}
\label{tpxy2a0p5}
\end{figure}

We show calculations of ${\cal B}_q(\{L\})$ in Figs. \ref{tpxy2a0p1}, 
\ref{tpxy2a0p2}, and \ref{tpxy2a0p5} for finite temperature.  
As $a$ increases from 0,
rather than having a fourfold degeneracy of $\lambda$'s at $q=3$, as in the
$a=0$ case, eq. (\ref{degeneq_q3}), one only has a twofold degeneracy, 
\beq
|\lambda_{L,5}|=|\lambda_{L,1}|
\label{degeneqqca}
\eeq
This occurs at the point 
\beq
q_c(\{L\})=q_c(\{ML\}) = \frac{(1-a)(3+2a)}{1+a} = 
-\frac{v(2v+5)}{v+2} \ . 
\label{qcl}
\eeq
Corresponding to this value of $q$ is the pair 
\beq
a_c(\{L\})_\pm=a_c(\{ML\})_\pm = 
\frac{1}{4}\biggl [ -(q+1) \pm \sqrt{q^2-6q+25} \ \biggr ] \ . 
\label{acl}
\eeq
As $a$ increases from 0 to 1, $q_c(\{L\})$ decreases monotonically from 3 to
0.  Over the interval $0 < a < a_d$, where 
\beq
a_d = \frac{1}{4}(-3+\sqrt{17} \ ) = 0.280776..
\label{am}
\eeq
the innermost region, $R_2$ continues to exist, but shrinks in size.  Its 
left-hand boundary on the real $q$ axis, where it is contiguous with region
$R_3$, continues to lie at 
\beq
q_{R_2-R_3} = 2 \quad {\rm for} \quad 0 < a < a_d \ . 
\label{qr2r3}
\eeq
This follows from the fact that this point is the real solution to the
degeneracy equation of leading terms $|\lambda_{L,1}|=|\lambda_{L,3}|$, and 
these are both equal (to $v^2$) at $q=2$.  As $a$
increases through the value $a=a_d$, $q_c(\{L\})_{a=a_d} = 2$, 
so this innermost region $R_2$ disappears; this 
can be seen from the fact that its right-hand boundary point $q_c$ becomes 
equal to its left-hand boundary point, given by $q_{R_2-R_3}$. 
Thus, for $a_d < a < 1$, the locus ${\cal B}_q$ separates the 
$q$ plane into only two regions, $R_1$ and $R_3$.

The Potts antiferromagnet has a phase transition at
the temperature given by the upper choice of the sign in eq. (\ref{acl}), i.e.,
\beq
T_{p,L} = \frac{J}{k_B\ln(a_c(\{L\})_+)} \ ,  \quad {\rm} \quad 0 < q < 3
\label{tpl}
\eeq 
(where both $J$ and the log are negative).  However, just as was found for
the analogous phase transition of the Potts model on the infinite-length
$L_y=2$ strip of the square lattice (which occurred for the range $0 < q < 2$
except for $f_{nq}$ for the integral value $q=1$) \cite{a}, the present phase
transition involves unphysical properties, including negative $Z$, negative
specific heat in the low-temperature phase, and non-existence of a
thermodynamic limit independent of boundary conditions in the low-temperature
phase.  The general existence of such pathologies was noted in \cite{ss}. 
As $a$ increases further to 1, ${\cal B}_q$ contracts
in to a point at the origin, $q=0$.

\subsubsection{Ferromagnetic range $a \ge 1$}

\begin{figure}[hbtp]
\centering
\leavevmode
\epsfxsize=2.5in
\begin{center}
\leavevmode
\epsffile{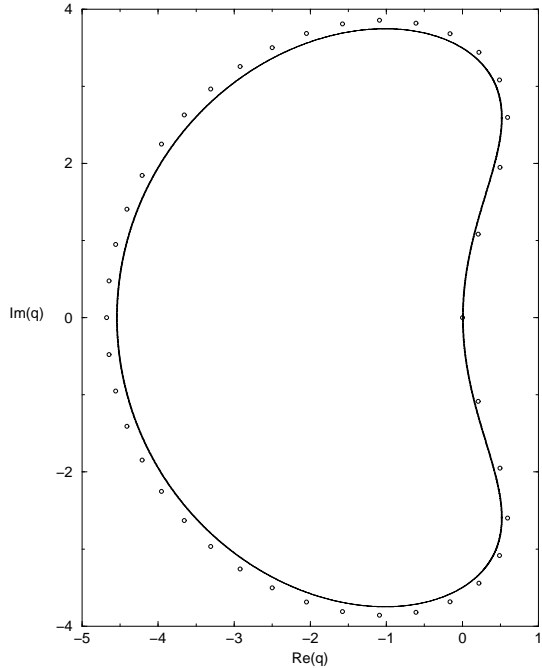}
\end{center}
\caption{\footnotesize{Locus ${\cal B}_q$ for the $n \to \infty$ limit of the
cyclic or M\"obius triangular strip, $\{L\}$ or $\{ML\}$, with $a=2$
(finite-temperature Potts ferromagnet).}}
\label{tpxy2a2}
\end{figure}

For the Potts ferromagnet, as $T$ decreases from infinity, i.e. $a$ increases
above 1, the locus ${\cal B}_q$ forms a lima-bean shaped curve shown for a
typical value, $a=2$, in Fig. \ref{tpxy2a2}.  Besides the generally present
crossing at $q=0$, the point $q_c(\{L\})$ at which ${\cal B}_q$ crosses the
real $q$ axis now occurs at negative $q$ values.  As was true of the model on
the analogous width $L_y=2$ cyclic and M\"obius strips of the square lattice
\cite{a}, for physical temperatures, the locus ${\cal B}_q$ for the Potts 
ferromagnet does not cross the positive real $q$ axis.  Note that 
this locus does have some support in the $Re(q) > 0$ half plane, away from the
real axis, which was also true of the analogous locus for the $L_y=2$ cyclic
and M\"obius square strip.

\subsection{${\cal B}_u(\{L\})$ for Fixed $q$}

We next proceed to the slices of ${\cal B}$ in the plane defined by the
temperature Boltzmann variable $u$, for given values of $q$, starting with
large $q$. In the limit $q \to \infty$, the locus ${\cal B}_u(\{L\})$ is
reduced to $\emptyset$.  This follows because for large $q$, there is only a
single dominant $\lambda_j$, namely
\beq
\lambda_{L,5} \sim q^2 + 4qv + O(1) \quad {\rm as } \quad q \to \infty \ .
\label{lambdar1asymp}
\eeq
Note that in this case, one gets the
same result whether one takes $q \to \infty$ first and then
$n=2m \to \infty$, or $n \to \infty$ and then $q \to \infty$, so that these
limits commute as regards the determination of ${\cal B}_u$.

We first consider values of $q \ne 0,1,2$, so that no noncommutativity occurs,
and $({\cal B}_u)_{nq}=({\cal B}_u)_{qn} \equiv {\cal B}_u$. As discussed
above, it is convenient to use the $u$ plane since ${\cal B}_u$ is compact in
this plane, except for the cases $q=2$ and $q=3$, whereas ${\cal B}_u$ is
noncompact because of the antiferromagnetic zero-temperature critical point at
$a=1/u=0$.

Extending the discussion in \cite{a} to the case of the strip of the 
triangular lattice, we observe that the property that the singular locus 
${\cal B}_u$ passes through the $T=0$ point $u=0$ for the Potts model with 
periodic boundary but not with free boundary conditions means that the use of
periodic boundary conditions yields a singular locus that manifestly
incorporates the zero-temperature critical point, while this is not manifest in
${\cal B}_u$ when calculated using free boundary conditions.  

\begin{figure}[hbtp]
\vspace{-20mm}
\centering
\leavevmode
\epsfxsize=2.5in
\begin{center}
\leavevmode
\epsffile{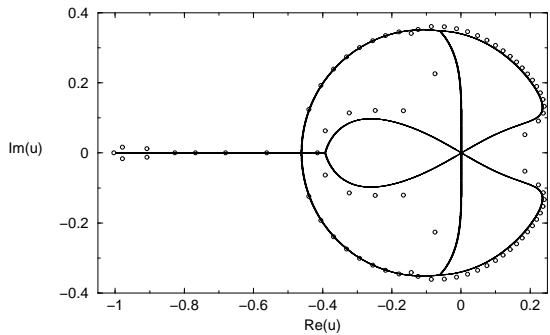}
\end{center}
\vspace{-20mm}
\caption{\footnotesize{Locus ${\cal B}_u(\{L\})$ for the $n \to \infty$ limit
of the cyclic or M\"obius triangular strip, $\{L\}$ or $\{ML\}$, with $q=10$.
Partition function zeros are shown for $m=20$, so that $Z$ is a polynomial of
degree $e=4m=80$ in $v$ and hence, up to an overall factor, in $u$).}}
\label{tpxy2q10}
\end{figure}

For $q=10$, the locus ${\cal B}_u$ is shown in Fig. \ref{tpxy2q10}.  Six curves
on ${\cal B}_u$ run into the origin, $u=0$, at the angles given in general in
eq. (\ref{thetau}).  The six corresponding complex-temperature phases
contiguous to the origin, $u=0$ exhaust the totality of such phases; i.e.,
there are no such phases that are disjoint from $u=0$.  The $\lambda_{L,j}$'s
that are dominant in these phases were given above, after eq. (\ref{thetau}).
As is evident in Fig. \ref{tpxy2q10}, part of ${\cal B}_u$ forms a
approximately circular curve.  The locus ${\cal B}_u$ also includes a line
segment on the negative real $u$ axis along which $\lambda_{L,5}$ and
$\lambda_{L,6}$ are dominant and are equal in magnitude as complex conjugates
of each other.

\begin{figure}[hbtp]
\centering
\leavevmode
\epsfxsize=2.5in
\begin{center}
\leavevmode
\epsffile{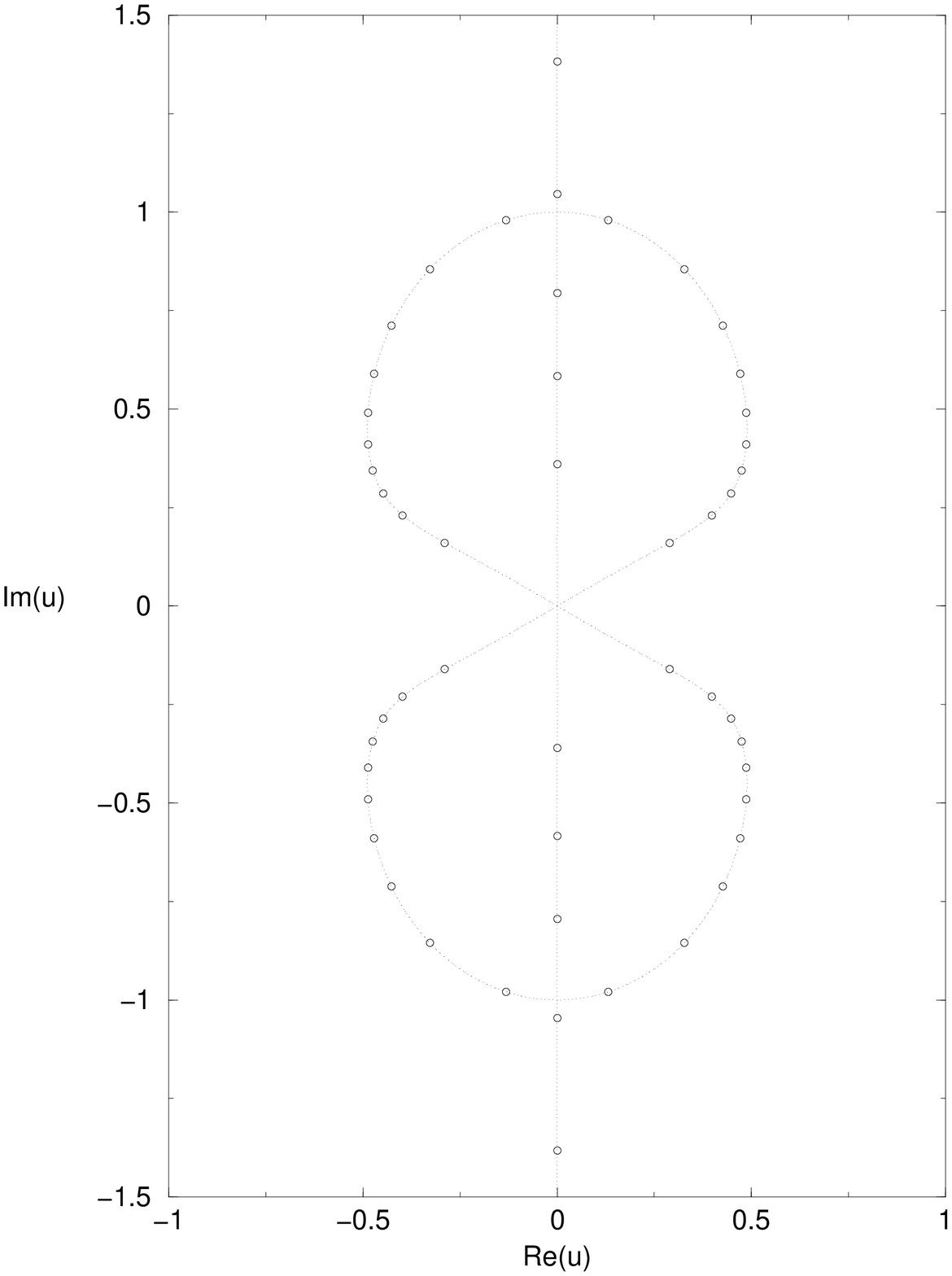}
\end{center}
\caption{\footnotesize{Locus ${\cal B}_u(\{L\})$: same as Fig. \ref{tpxy2q10}
for $q=2$.}}
\label{tpxy2q2}
\end{figure}

For $q=2$, the locus $({\cal B}_u)_{nq}$ is shown in Fig. \ref{tpxy2q2}.  One
sees that, in addition to the six curves intersecting at the ferromagnetic
zero-temperature critical point $u=0$, ${\cal B}_u$ has multiple points at
$u=\pm i$ where four curves intersect.  The complex-temperature phases in the
vicinity of $u=0$ were determined above after eq. (\ref{thetau}), and these
exhaust all complex-temperature phases, i.e. there is none that does not extend
in to $u=0$. 

For $q=3$, the locus ${\cal B}_u$ was given as Fig. 10 in Ref. \cite{1dnnn} and
is shown with associated partition function zeros in Fig. \ref{tpxy2q3}. In
this case, in addition to the six phases that are contiguous at $u=0$, there is
evidently another $O$ phase that includes the negative real axis for $u < -1$,
and a complex conjugate pair of $O$ phases extending outward from the
intersection points at $u=e^{\pm 2 \pi i/3}$ toward the upper and lower left.
At these intersection (multiple) points, six curves on ${\cal B}_u$ intersect,
just as was true at $u=0$.  The existence of the intersection points on ${\cal
B}_u$ at $u=e^{\pm 2 \pi i/3}$ for the strip would suggest that such points
could also be present on the locus ${\cal B}_u$ for the $q=3$ Potts model on
the full triangular lattice.  Since this model has not been exactly solved, one
can only try to infer ${\cal B}_u$ from complex-temperature (Fisher) zeros of
the partition function calculated for large triangular lattices \cite{mm,p,p2}.
These are consistent with this possibility (see, e.g., Fig. 1 of \cite{mm}
or Figs. 5-7 of \cite{p2}) although the zeros show such a high degree of
scatter in the $Re(a) < 0$ half-plane that one cannot draw a very decisive
conclusion from them.  

For $q=4$ we show the locus ${\cal B}_u$ in Fig. \ref{tpxy2q4}.  In this case,
we remark, in

\begin{figure}[hbtp]
\centering
\leavevmode
\epsfxsize=2.5in
\begin{center}
\leavevmode
\epsffile{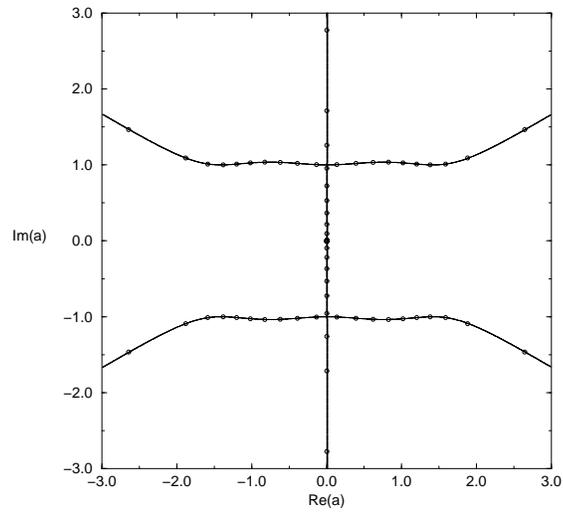}
\end{center}
\vspace{-10mm}
\caption{\footnotesize{Locus ${\cal B}_a$ for the $n \to \infty$ limit of the
cyclic or M\"obius triangular strip, $\{L\}$ or $\{ML\}$, with $q=2$.}}
\label{tpxy2aq2}
\end{figure}

\begin{figure}[hbtp]
\centering
\leavevmode
\epsfxsize=2.5in
\begin{center}
\leavevmode
\epsffile{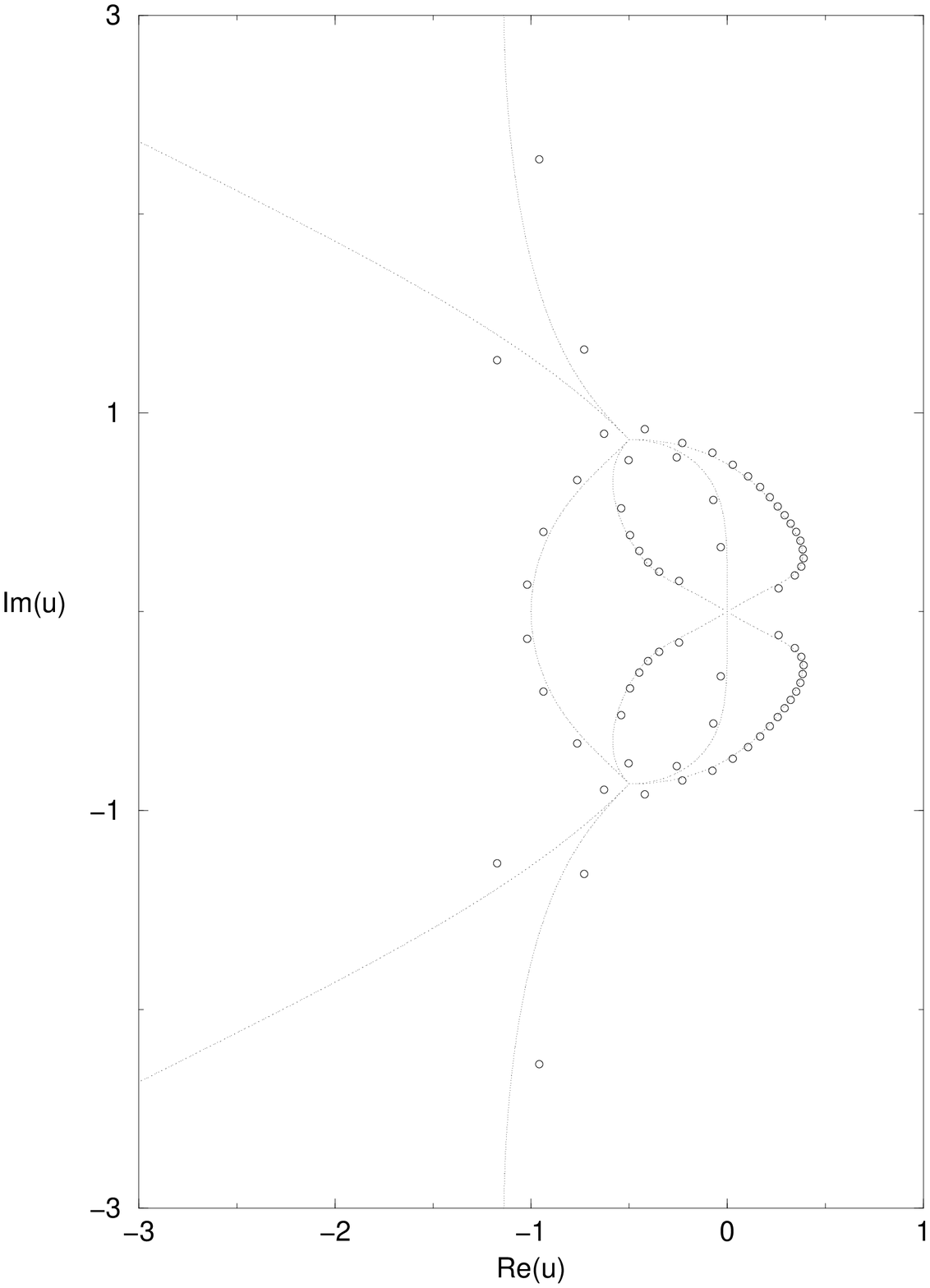}
\end{center}
\caption{\footnotesize{Locus ${\cal B}_u(\{L\})$: same as Fig. \ref{tpxy2q10}
for $q=3$.}}
\label{tpxy2q3}
\end{figure}

\begin{figure}[hbtp]
\centering
\leavevmode
\epsfxsize=2.5in
\begin{center}
\leavevmode
\epsffile{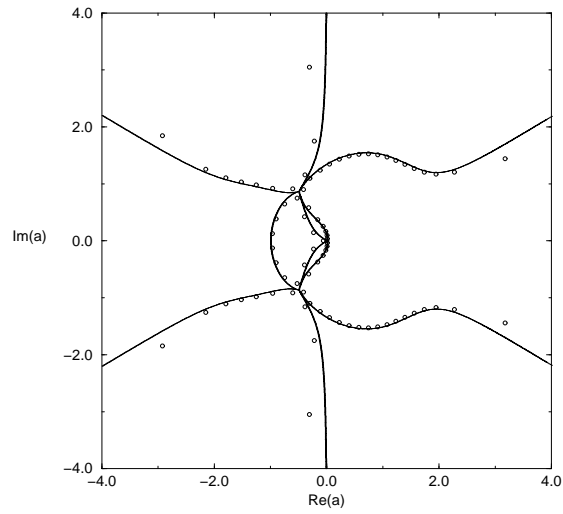}
\end{center}
\vspace{-10mm}
\caption{\footnotesize{Same as Fig. \ref{tpxy2aq2} for $q=3$.}}
\label{tpxy2aq3}
\end{figure}

\begin{figure}[hbtp]
\vspace{-20mm}
\centering
\leavevmode
\epsfxsize=2.5in
\begin{center}
\leavevmode
\epsffile{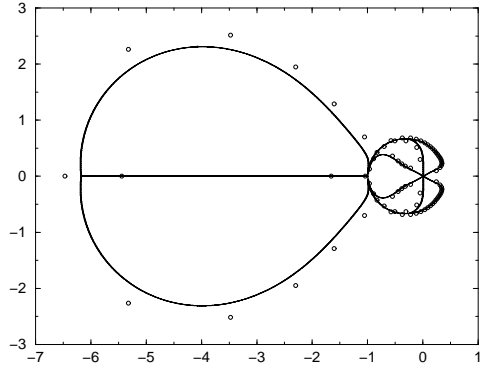}
\end{center}
\vspace{-20mm}
\caption{\footnotesize{Same as Fig. \ref{tpxy2q2} for $q=4$.}}
\label{tpxy2q4}
\end{figure}

For the study of the zero-temperature critical point of the Potts
antiferromagnet on the $L_y=2$ cyclic and M\"obius strips of the triangular
lattice, it is useful to display the singular locus ${\cal B}_a$ in the $a$
plane, since the critical point occurs at $a=0$.  We show these plots for $q=2$
and 3 in Figs. \ref{tpxy2aq2} and \ref{tpxy2aq3}.  For $q=2$, the boundary
${\cal B}_a$ has a multiple point at $a=\pm i$ where four branches intersect.
For $q=3$, four curves on ${\cal B}_a$ pass through $a=0$ at the angles
given in eqs. (\ref{phi1}) and (\ref{phi2}).  Note that that the curves that 
leave the origin at the angles $\phi$ given in eq. (\ref{phi1}) rapidly bend 
toward the vertical and then back toward the left, so that the
complex-temperature phase that is contiguous with $u=0$ and lies in the 
angular wedges between $76^\circ$ and $150^\circ$, and its complex-conjugate,
are rather narrow.  The multiple points at $a=e^{\pm 2\pi i/3}$ have
been described above.

\begin{figure}[hbtp]
\centering
\leavevmode
\epsfxsize=2.5in
\begin{center}
\leavevmode
\epsffile{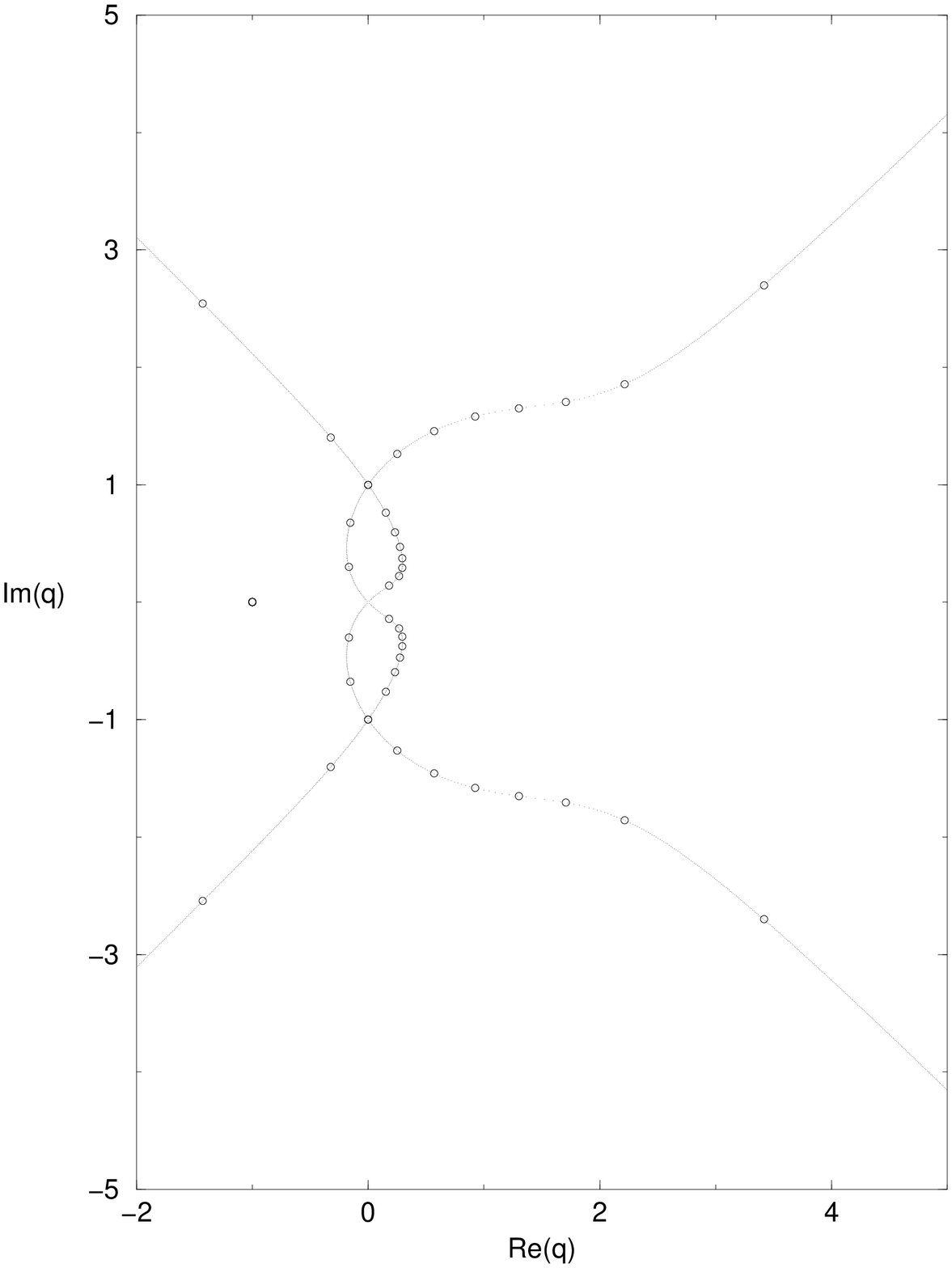}
\end{center}
\caption{\footnotesize{Locus ${\cal B}_u={\cal B}_a$ for the $L_y=2$ square 
strip with $q=2$. Partition function zeros are shown for $m=20$ ($e=60$).}}
\label{sqpxy2q2}
\end{figure}

\begin{figure}[hbtp]
\centering
\leavevmode
\epsfxsize=2.5in
\begin{center}
\leavevmode
\epsffile{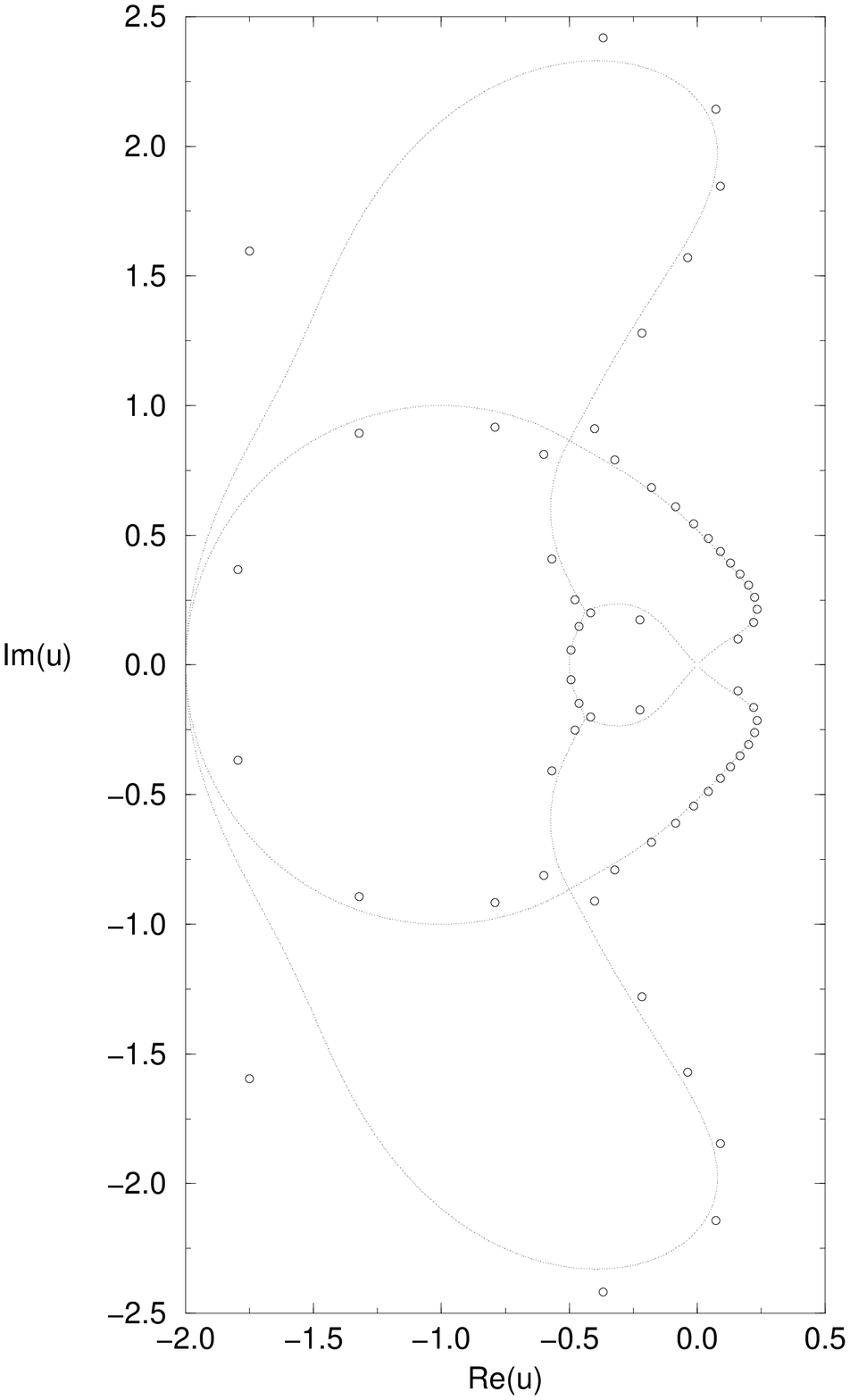}
\end{center}
\caption{\footnotesize{Locus ${\cal B}_u$ for the $L_y=2$ square strip with 
$q=3$. Partition function zeros are shown for $m=20$ (so that $Z$ is a
polynomial of degree $e=3m=60$ in $v$ and hence, up to an overall factor, in
$u$).}}
\label{sqpxy2q3}
\end{figure}

In Figs. \ref{sqpxy2q3} and \ref{sqpxy2q2} we show the analogous loci 
${\cal B}_u$ for the $L_y=2$ cyclic or M\"obius strip of the square lattice
studied in \cite{a} for comparison.  A particularly striking difference is
that, since the $q=3$ Potts antiferromagnet is (is not) critical on the $L_y=2$
strip of the triangular (square) lattice, the resultant locus ${\cal B}_u$
passes through (does not pass through) $1/u=a=0$, respectively. 
The Ising model, $q=2$ has both ferromagnetic and antiferromagnetic $T=0$ 
critical points on both the $L_y=2$ square and triangular lattice strips, and 
hence for both strips, ${\cal B}_u$ passes through $1/u=a=0$.

\begin{figure}[hbtp]
\centering
\leavevmode
\epsfxsize=2.5in
\begin{center}
\leavevmode
\epsffile{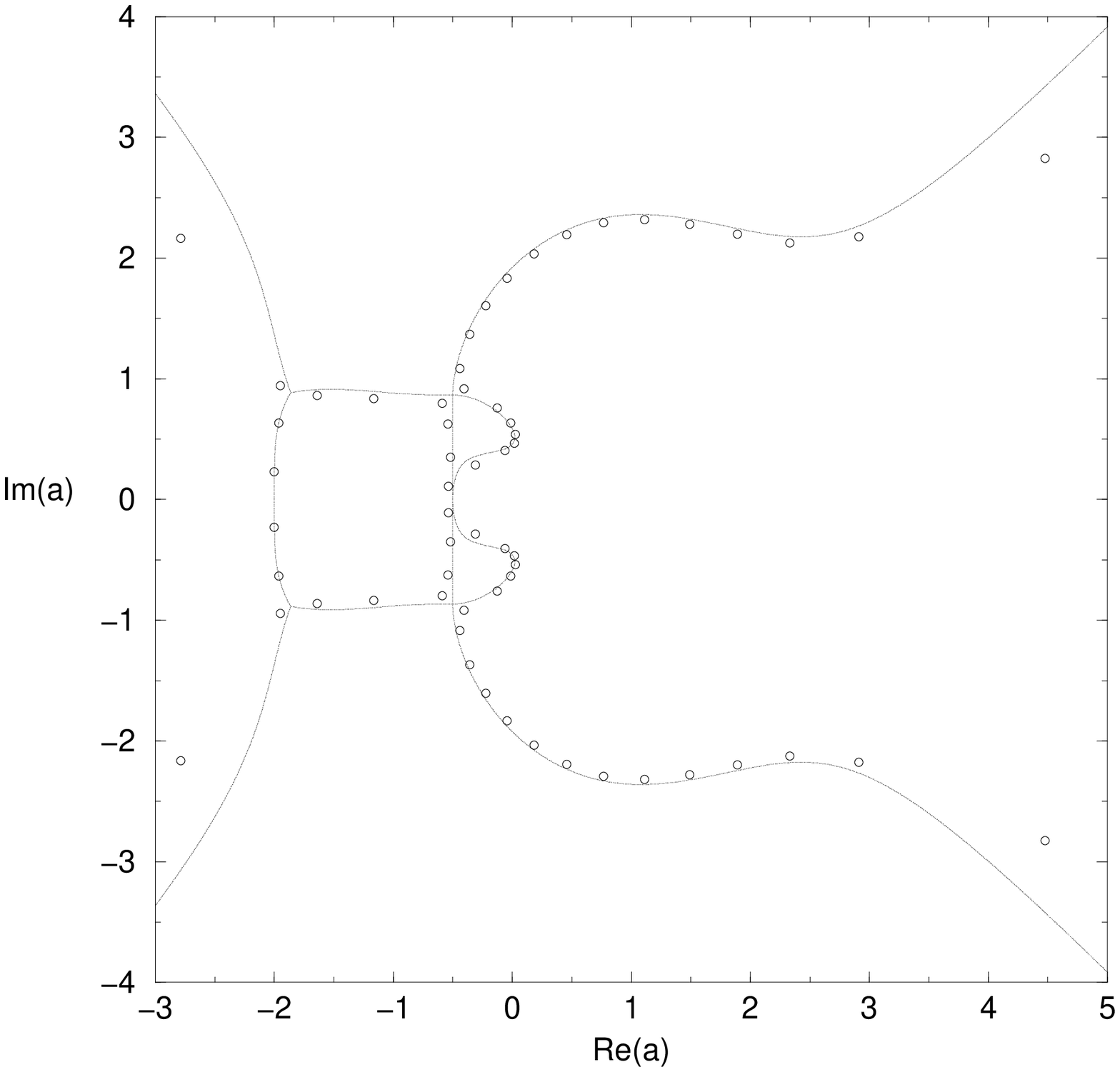}
\end{center}
\vspace{-10mm}
\caption{\footnotesize{Locus ${\cal B}_a$ for the $L_y=2$ square strip with
$q=3$. Partition function zeros are shown for $m=20$ (so that $Z$ is a
polynomial of degree $e=3m=60$ in $v$ and hence $a$).}}
\label{sqpxy2aq3}
\end{figure}

\subsection{Connections Between ${\cal B}$ for Strips and 2D Lattices}

In earlier work \cite{hsi,is1d,a}, it was shown that although the physical
thermodynamic properties of a discrete spin model are, in general, different in
1D or infinite-length, finite-width strips, which are quasi-1D systems, and in
higher dimensions, nevertheless, exact solutions for ${\cal B}_u$ in 1D and
quasi-1D systems can give insight into ${\cal B}_u$ in 2D.  This was shown, in
particular, for the $q=2$ Ising special case of the Potts model, where the
comparison can be made rigorously since the model is exactly solvable in 2D. 
It will often be convenient below to use the equivalent locus in the $a=1/u$
plane, ${\cal B}_a$. 
In \cite{wn,strip,w2d,bcc,wcy} it was also noted how exact calculations of 
${\cal B}_q$ and $W$ on infinite-length, finite-width strips can give
information about the behavior of these quantities on 2D lattices.  

In Ref. \cite{a}, the connection between ${\cal B}_a$ on strips and in on the
full 2D square lattice was studied; here we shall do this for the triangular
lattice.  Again, it is natural to start with the $q=2$ Ising case, where one
has exact results both for the strip and the full triangular lattice.  The
comparisons made in \cite{a} and here enable one to formulate a reasonably
systematic procedure for making transformations on an exactly calculated locus
${\cal B}_a$ for the Potts model on an infinite-length, finite-width strip of a
given lattice, in order to construct at least a qualitatively correct locus
${\cal B}_a$ on the corresponding 2D lattice.  As emphasized in \cite{a}, this
represents a new and powerful way of obtaining information about
complex-temperature phase diagrams of spin models in 2D (indeed perhaps also
higher dimensions) from exact results on strips.  It will be recalled that the
conventional approach has been the rather laborious procedure of performing
exact calculations of the partition function on finite 2D lattices of various
sizes with various boundary conditions \cite{mm}-\cite{pfef}.  Although
some features of ${\cal B}_a$ for the Potts model, such as the circle
$|a-1|=\sqrt{q}$ for $Re(a) > 0$ are evident with this conventional procedure,
the comlex-temperature (Fisher) zeros often exhibit such considerable scatter
for $Re(a) < 0$ that it is difficult to infer the structure of the locus ${\cal
B}_a$ in this region.  One method was to combine calculations of the zeros with
dlog Pad\'e and differential approximant analyses of low-temperature series
expansions so as to localize accurately certain points on the
complex-temperature boundary \cite{pfef}-\cite{p2}.  This method was motivated
by the fact that complex-temperature singularities in such quantities as 
specific heat and magnetization for the 2D Ising model can be calculated
exactly and they occur at known points on the complex-temperature boundaries
for respective 2D lattices; it was also shown that this was true for the
susceptibility, where no exact calculation exists \cite{chisq,chitri}.
However, while this method avoids the problem of the scatter in Fisher zeros,
it only localizes certain points on ${\cal B}_a$.  The present method is
complementary in that it enables one to gain, at least qualitatively, an idea
of the global structure of ${\cal B}_a$. 

We proceed with our exact comparison for the $q=2$ Ising case and start by
recalling the situation for the square lattice \cite{a}.  The locus ${\cal
B}_u$ for the square lattice consists of the well-known union of circles
$|u-1|=\sqrt{2}$ and $|u+1|=\sqrt{2}$.\footnote{\footnotesize{In Refs.
\cite{chisq,chitri} the variable $z=e^{-2K_{Ising}} \equiv e^{-K_{Potts}}$ was
used for what we denote as $u$ here and the variable $u$ was used for
$e^{-4K_{Ising}} \equiv e^{-2K_{Potts}}$, i.e. for what would be denoted $u^2$
here. Thus, the circles $|u \pm 1|=\sqrt{2}$ map to the lima\c{c}on of Pascal
in the $u^2$ plane \cite{chisq}.}}  The procedure for transforming the locus
${\cal B}_u$ found for the infinite-length, width $L_y=2$ strip in \cite{a}
(see Fig. \ref{sqpxy2q2}) involves two main steps.  First, one retracts each of
the curves going through the origin $u=0$ so that they no longer pass through
this point.  It is clear that this retracting is necessary not just for $q=2$
but for the general $q$-state Potts ferromagnet since the (reduced) free energy
and magnetization of this model are analytic in the neighborhood of $u=0$,
i.e., they have low-temperature series expansions a finite radius of
convergence.  Given the inversion symmetry ${\cal B}_u={\cal B}_a$ that holds
for the square lattice and strips of this lattice (but not for the triangular
lattice), this retracting means that the curves are also pulled back from the
origin of the $a$ plane.  The second step in the procedure is to incorporate
the feature that the 2D Potts ferromagnet has a finite-temperature phase
transition.  To build in this feature, one takes the two complex-conjugate ends
of the curves with $Re(u)$ small and positive that have been pulled back from
the origin $u=0$ and connects them so that they cross the positive real $u$
axis at a point $u_{PM-FM}$ in the interval $0 < u < 1$; by the inversion
symmetry this has the effect of connecting the two other ends in the $Re(u) >
0$ half plane and having them cross the real axis at $u_{PM-AFM} =
u_{PM-FM}^{-1}$.  The third step is to build in the feature
\cite{chisq,chitri,cmo} that if a lattice has an even coordination number, then
the Ising model boundary ${\cal B}_u$ is symmetric under $u \to -u$; in the
present case, the locus ${\cal B}_u$ for the $L_y=2$ strip does not have this
property because the coordination number of the vertices on the strip is 3,
while the locus ${\cal B}_u$ for the square lattice does have the property.  We
thus connect the curves in the left-hand half plane $Re(u) < 0$ in such a way
as to have this property.  As discussed in \cite{a}, the intersection points at
$u=\pm i$ are the same for both the strip and the square lattice, and, indeed,
on wider strips.  Evidently, each step of this procedure is based on
fundamental principles, and, as long as one limits oneself to obtaining the
qualitative features of the locus ${\cal B}_u$ for the 2D case from the $L_y=2$
strip, nothing is {\it ad hoc}.  Of course, one cannot predict the actual
values of the critical point $u_{PM-FM}=u_{PM-AFM}^{-1}$, but this is was not
the goal; there are powerful ways of determining the critical point via series
analyses even in cases where a spin model cannot be solved exactly in 2D. (Note
that for the purpose of obtaining information about ${\cal B}_u$ for a model in
2D, it can be advantageous to use strips with width $L_y=2$ rather than wider
strips, since the locus ${\cal B}_u$ becomes progressively more complicated as
$L_y$ increases, with more curves passing through $u=0$ \cite{a}.)

Next, we demonstrate the corresponding comparison with the locus ${\cal B}_u$
for the $q=2$ Ising special case of our new results for the Potts model on the
$L_y=2$ strip of the triangular lattice and ${\cal B}_u$ for the model on the
full 2D triangular lattice.  Our calculations are shown in Figs. \ref{tpxy2q2}
and \ref{tpxy2aq2}.  For the Ising model on the full 2D triangular lattice,
${\cal B}_u$ consists of the union of an oval curve that crosses the real
(imaginary) $u$ axis at $\pm i$ with the line segment $1/\sqrt{3} \le Im(u) <
\infty$ and its complex conjugate on the imaginary $u$ axis \cite{chitri}.
Note that, just as was true in the case of the square strips and square lattice
(and the honeycomb lattice \cite{chitri}), the locus ${\cal B}_u$ for the Ising
model on both the current strip and the full 2D triangular lattice has
intersection points at $u = \pm i$.  (In terms of the variable denoted $u^2$ in
our current notation (and $u$ in the notation of \cite{chitri}), the Ising
locus for the triangular lattice is the union of the semi-infinite line segment
$-\infty < u^2 < -1$ and the circle $|u^2+(1/3)|=2/3$, and the above two
intersection points map to the single intersection point at $u^2=-1$.)
Following the same procedure as for the square strip, the first step is to pull
back each of the curves (of which there are now six rather than four) from the
origin.  On the imaginary axis, this retracting produces two complex-conjugate
semi-infinite line segments, while for the other four curves, it allows one to
connect them smoothly, via step 2, to form the oval.  Step 3 is not necessary
here since both the cyclic $L_y=2$ strip of the triangular lattice and the full
triangular lattice have the property that each vertex has even coordination
number ($\Delta=4$ for the strip and $\Delta=6$ for the 2D lattice) so that 
${\cal B}_u$ is invariant under the replacement $u \to -u$ \cite{chitri,cmo}. 

It is also interesting to observe that the similarities between the Ising model
on the current strip and on the full 2D lattice are much stronger for the
antiferromagnet than for the ferromagnet.  In contrast to the case of the Ising
ferromagnet, which has a finite-temperature critical point on 2D lattices but
only a zero-temperature critical point on 1D and quasi-1D lattices, the Ising
antiferromagnet has a zero-temperature critical point on both the full 2D
triangular lattice \cite{istri} and on the infinite-length, width $L_y=2$
strip of the triangular lattice, as a consequence of the frustration in this
model.  Thus in this case, there are not just similarities in the
complex-temperature properties of the model, but in the actual physical
thermodynamics also.  Concerning the complex-temperature phase diagram, in the
vicinity of this zero-temperature critical point, for the full 2D triangular
lattice, ${\cal B}_a$ is just the inverse image of the locus ${\cal B}_u$ given
above, namely the union of an oval intersecting the real (imaginary) $a$ axes
at $\pm \sqrt{3}$ and $\pm i$, and a finite line segment on the imaginary axis
stretching from $a=-\sqrt{3}i$ to $a=\sqrt{3}i$.  In the neighborhood of the
origin, $a=0$, the locus ${\cal B}_a$ for the present strip is exactly the
same, as can be seen in Fig. \ref{tpxy2aq2}. 

Having shown the correspondences between the complex-temperature boundaries
${\cal B}_u$ and the associated complex-temperature phase diagrams for the
exactly solved Ising case, we now use this result as a tool to suggest features
of ${\cal B}$ for the 2D Potts model for other values of $q$, where it has not
been exactly solved.  Since the complex-temperature zeros have usually been
presented in the $a$ plane, we follow this convention here.  The case $q=3$ on
strips of the square lattice and on the full square lattice was discussed in
\cite{a}.  In Figs. \ref{sqpxy2q3} and \ref{sqpxy2aq3} we show ${\cal B}_u$
and ${\cal B}_a$ for the $L_y=2$ strip of the square lattice.  First, the fact
that ${\cal B}_a$ crosses the real axis at $a=-2$ for the strip is the same as
is true for the square lattice \cite{a}; in the latter case, this follows from
the existence of the zero-temperature critical point for the $q=3$
antiferromagnet, the resultant fact that the singular locus ${\cal B}_a$ passes
through $a=0$, and the duality of the model, according to which if a point $a$
is on ${\cal B}_a$, then the dual image $a_d=(q+a-1)/(a-1)$ is also on ${\cal
B}_a$.  Second, the existence of the intersection points at $a=e^{\pm 2\pi
i/3}$ on ${\cal B}_a$ for the strip suggests that these also occur on ${\cal
B}_a$ for the square lattice.  We now extend these two comments to explore the
global structure of ${\cal B}_a$.  Since we know that the 2D $q=3$ Potts model
on the square lattice is analytic in the neighborhood of the $T=0$
ferromagnetic point $u=0$, our first step is to pull back the four curves that
pass through $u=0$ in Fig. \ref{sqpxy2q3} (equivalently, off to infinity in
Fig. \ref{sqpxy2aq3}).  Since we know that the 2D model has a
finite-temperature critical point (the exact value, $a=1|sqrt{3}$ is not
crucial here), we connect the two ends of the curves in the $Re(a) > 0$
half-plane together in Fig. \ref{sqpxy2aq3} so that they cross the positive
real $a$ plane, making the ferromagnetic critical point $a_{PM-FM} > 1$.  The
ends of the two curves pointing in the upper left and lower left directions in
the $Re(a) < 0$ half-plane are left as is, in accordance with the finding from
series analyses \cite{pfef} that there are such singularities at $a_e =
-1.71(1) \pm 1.46(q)i$, consistent with lying on endpoints of curves on ${\cal
B}_a$.  Since we know that the $q=3$ Potts antiferromagnet has a $T=0$ critical
point on the square lattice, and hence ${\cal B}_a$ passes through $a=0$, we
move the four curves that pass through the point $a=-1/2$ (forming an
intersection point) over to the right, so that at least one branch passes
through $a=0$; the exact solution for the strip suggests that ${\cal B}_a$ for
the square lattice may also have an intersection (multiple) point at $a=0$.
The complex-temperature zeros for the $q=3$ Potts model that have been
calculated for patches of the square lattice \cite{mm,mbook,wuetal,pfef} are
consistent with this suggestion (see, e.g., Fig. 1 of \cite{pfef}).  A third
suggestion is that the intersection points on ${\cal B}_a$ at $a \simeq -1.85
\pm 0.85i$ for the strip have analogues on the locus ${\cal B}_a$ for the full
square lattice.  This suggestion is consistent with the patterns of
complex-temperature zeros that have been calculated for patches of the square
lattice, but the scatter is too great to provide a strong test. 

We next use our new results on the $q=3$ Potts model on the $L_y=2$ strip of
the triangular lattice. 

\subsection{Thermodynamics of the Potts Model on the $L_y=2$ Strip of the
Triangular Lattice}

\subsubsection{Ferromagnetic Case}

\begin{figure}[hbtp]
\centering
\leavevmode
\epsfxsize=2.5in
\begin{center}
\leavevmode
\epsffile{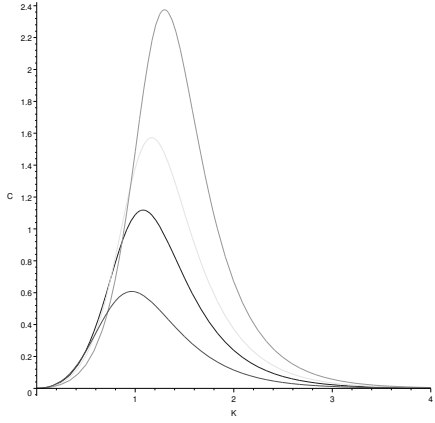}
\end{center}
\vspace{-25mm}
\caption{\footnotesize{Specific heat (with $k_B \equiv 1$) for the Potts
ferromagnet on the infinite-length, width $L_y=2$ strip of the triangular
lattice, as a function of $K=J/(k_BT)$.  Going from bottom to top in order of
the heights of the maxima, the curves are for $q=2,3,4,6$.}}
\label{cfmtristrip}
\end{figure}

The Potts ferromagnet (with real $q > 0$) on an arbitrary graph has $v > 0$ so,
as is clear from eq. (\ref{cluster}), the partition function satisfies the
constraint of positivity.  In contrast, the specific heat $C$ is positive for
the model on the (infinite-length limit of the) $L_y=2$ triangular strip is
positive if and only if $q > 1$ (for any choice of longitudinal boundary
conditions). For $q=1$, $f_{nq}=2\ln a = 2K$ and $C$ vanishes identically.
Since a negative specific heat is unphysical, we therefore restrict to real $q
\ge 1$. For general $q$ in this range, the reduced free energy is given for all
temperatures by $f=(1/2)\ln \lambda_{S,1}$ as in (\ref{fstrip}) (independent of
the different longitudinal boundary conditions, as must be true for the
thermodynamic limit to exist).  Recall that $\lambda_{S,1} \equiv
\lambda_{L,5}$.  It is straightforward to obtain the internal energy $U$ and
specific heat from this free energy; since the expressions are somewhat
complicated, we do not list them here.  We show a plot of the specific heat
(with $k_B=1$) in Fig. \ref{cfmtristrip}.  One can observe that the value of
the maximum is a monotonically increasing function of $q$.

The high-temperature expansion of $U$ is
\beq
U=-\frac{2J}{q}\biggl [ 1+\frac{(q-1)}{q}K + O(K^2) \biggr ] \ .
\label{ustriphigh}
\eeq
For the specific heat we have
\beq
C=\frac{2k_B(q-1)K^2}{q^2}\biggl [ 1 + \frac{(q+1)}{q}K + O(K^2) \biggr ] \ . 
\label{cstriphigh}
\eeq
The low-temperature expansions ($K \to \infty$) are 
\beq
U = J\Biggl [ -2 + (q-1)e^{-3K}\biggl [ 3 + 4e^{-K} + 5(q-1)e^{-2K} 
+ O(e^{-3K}) \biggr ] \Biggr ] \quad {\rm as} \quad K \to \infty
\label{ustriplowfm}
\eeq
and 
\beq
C = 9k_BK^2(q-1)e^{-3K}\biggl [1+\frac{16}{9}e^{-K} + 
\frac{25}{9}(q-1)e^{-2K} + O(e^{-3K})\biggr ] \quad {\rm as} \quad K \to \infty
\label{cstriplowfm}
\eeq

In general, the ratio $\rho$ of the largest subdominant to the dominant
$\lambda_j$'s determines the asymptotic decay of the connected spin-spin
correlation function and hence the correlation length 
\beq
\xi = -\frac{1}{\ln \rho}
\label{xi}
\eeq
Since $\lambda_{L,5}$ and $\lambda_{L,2}$ are the dominant and leading
subdominant $\lambda_j$'s, respectively, we have 
\beq
\rho_{FM}=\frac{\lambda_{L,2}}{\lambda_{L,5}}
\label{rho}
\eeq
and hence for the ferromagnetic zero-temperature critical point we find that 
the correlation length diverges, as $T \to 0$, as 
\beq
\xi_{FM} \sim (2q)^{-1}e^{3K} \ , \quad {\rm as} \quad K \to \infty
\label{xit}
\eeq

\subsubsection{Antiferromagnetic Case}

In this section we first restrict to the real range $q \ge 3$ and the
additional integer value $q=2$ (Ising case) where the Potts 
antiferromagnet exhibits physically acceptable behavior and then consider the
remaining interval $0 < q < 3$ where (except for the trivial $f_{nq}$ for
$q=1$) it exhibits unphysical properties.  For $q
\ge 3$, the free energy is given for all temperatures by (\ref{fstrip}), as in
the ferromagnetic case but with $J$ negative, and is the same independent of
the different longitudinal boundary conditions, as is necessary for there to
exist a thermodynamic limit. 

\begin{figure}[hbtp]
\centering
\leavevmode
\epsfxsize=2.5in
\begin{center}
\leavevmode
\epsffile{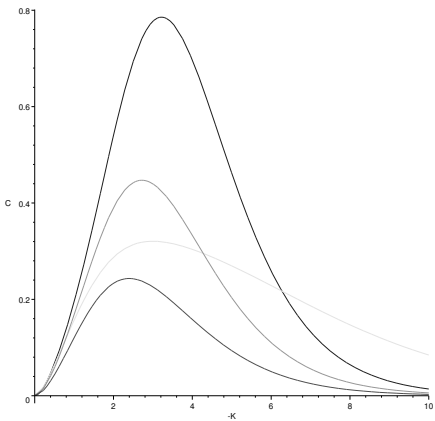}
\end{center}
\vspace{-25mm}
\caption{\footnotesize{Specific heat (with $k_B \equiv 1$) for the Potts
antiferromagnet on the infinite-length, width $L_y=2$ strip of the triangular
lattice, as a function of $-K = -J/(k_BT)$.  Going downward in order of the
heights of the maxima, the curves are for $q=3,4,2,6$.}}
\label{cafmtristrip}
\end{figure}

We show plots of the specific heat, for several values of $q$, for the Potts
antiferromagnet on the (infinite-length limit of the) $L_y=2$ strip of the
triangular lattice in Fig \ref{cafmtristrip}.  In contrast to the ferromagnetic
case, the maxima of $C$ do not increase monotonically with $q$; they occur at
0.32 for $q=2$ and 0.79 for $q=3$, after which the values of the maxima
decrease with increasing integral $q$ (e.g., they are 0.45 and 0.24 for $q=4$
and $q=6$).  The fact that the curve for $C$ for the $q=2$ (Ising)
antiferromagnet exhibits a very broad maximum can be inferred to be a
consequence of the frustration that is present in this model.

The high-temperature expansions of $U$ and $C$ are given by (\ref{ustriphigh})
and (\ref{cstriphigh}); more generally, these expansions also apply in the
range $0 < q < 3$.  As discussed above, the Ising case $q=2$ is one of the
cases where one must take account of noncommutativity in the definition of the
free energy and hence of thermodynamic quantities.  If one sets $q=2$ first and
then takes $n \to \infty$, then $f=f_{nq}=(1/2)\ln \lambda_{L,5}|_{q=2}$ where 
$\lambda_{L,5}|{q=2}$ was given in eq. (\ref{lam56q2}), and the 
low-temperature expansions are 
\beq
U_{q=2}=-\frac{J}{2}\biggl [ 1 + \frac{1}{2}e^{K/2} + \frac{23}{16}e^{3K/2} - 
2e^{2K} + O(e^{5K/2}) \biggr ] \quad {\rm as} \quad K \to -\infty
\label{ustriplowafmising}
\eeq
and
\beq
C_{q=2} = \frac{k_BK^2e^{K/2}}{8}\biggl [ 1 + \frac{69}{8}e^K - 16e^{3K/2} +
\frac{1215}{128}e^{2K} + O(e^{3K}) \biggr ] 
\quad {\rm as} \quad K \to -\infty \ . 
\label{cstriplowfmising}
\eeq

For the range $q \ge 3$, the low-temperature expansions are given by 
\beq U =
\frac{(-J)e^K}{(q-2)^2}\biggl [ (2q-3) - \frac{(2q-3)(4q-5)}{(q-2)^2} e^K +
O(e^{2K}) \biggr ] \quad {\rm as} \quad K \to -\infty
\label{ustriplowafm}
\eeq
and
\beq
C = \frac{k_BK^2e^K}{(q-2)^2}\biggl [ (2q-3) -
\frac{2(2q-3)(4q-5)}{(q-2)^2} e^K + O(e^{2K}) \biggr ]
\quad {\rm as} \quad K \to -\infty \ .
\label{cstriplowafm}
\eeq 
Note that for the antiferromagnetic case, $U(T=0)=0$ for $q \ge 3$, but
$U(T=0)=-J/2=+|J|/2$ for $q=2$.  The vanishing value of $U$ at $T=0$ for $q \ge
3$ means that the Potts model can achieve its preferred ground
state for this range of $q$, while the nonzero value of $U(T=0)$ for the Ising
antiferromagnet is a consequence of the frustration that is present in this
case.  Similarly, the fact that the specific heat vanishes less rapidly for the
Ising antiferromagnet than for the Potts model with $q \ge 3$ reflects the
frustration that is present in the Ising case, which is not present for $q \ge
3$. Note that the apparent divergences that occur as $q \to 2$ in
eqs. (\ref{ustriplowafm}) and (\ref{cstriplowafm}) are not actually
reached here since these expressions apply only in the region $q \ge 3$ (the
discrete integral case $q=2$ was dealt with above).

For the zero-temperature critical points in the $q=2$ and $q=3$ Potts
antiferromagnet, 
\beq
\rho_{AFM,q=2,3}=\frac{\lambda_{L,2}}{\lambda_{L,5}}
\label{rhoafmq2}
\eeq
Using the respective expansions (\ref{lam24q2taylor})-(\ref{lam56q2taylor}) and
(\ref{lam23q3taylor})-(\ref{lam6q3taylor}), we find
that the correlation lengths defined as in (\ref{xi}) diverges, as
$T \to 0$, as
\beq
\xi_{AFM,q=2} \sim e^{-K/2} \ , \quad {\rm as} \quad K \to -\infty
\label{xiafmq2}
\eeq
and
\beq
\xi_{AFM,q=3} \sim e^{-K} \ , \quad {\rm as} \quad K \to -\infty
\label{xiafmq3}
\eeq

Next, we consider the range of real nonintegral $0 < q < 3$.  The first
pathology is that the Potts antiferromagnet on the infinite-length limit of the
$L_y=2$ triangular strip, defined with case of cyclic or M\"obius longitudinal
boundary conditions has a phase transition at the temperature $T_{p,L}$ given
in eq. (\ref{tpl}) for $0 < q < 3$ (except for $f_{nq}$ for the integral
values$q=1,2$), while, in contrast, if one uses free boundary conditions, then
there is no phase transition at this temperature and, although there is a phase
transition for $0 < q < 2$, it occurs at the temperature $T_{p,S}$ given in
eq. (\ref{tpstrip}), which, in general, is not equal to $T_{p,L}$.  It follows
that there is no well-defined thermodynamic limit for the Potts model 
with $0 < q < 3$ and $q \ne 1,2$.  The Ising case $q=2$ has been dealt with in
the preceding subsection.  Concerning the value $q=1$, as discussed earlier,
one encounters noncommutativity in defining the free energy.  If one takes
$q=1$ to start with and then $n \to \infty$, the thermodynamic limit does
exist, independent of boundary conditions, and $f=f_{nq}=2K$, $U=-2J=2|J|$, and
$C=0$.  If one starts with $q \ne 1$, takes $n \to \infty$, calculates
$f_{qn}$, and then takes $q \to 1$, the thermodynamic limit does not exist
since the result differs depending on whether one uses free longitudinal
boundary conditions or cyclic (equivalently M\"obius) longitudinal boundary
conditions.  Specifically, for this single value $q=1$, $T_{p,S}$ is equal to
$T_{p,L}$, having the value given by $k_BT_{p,S}=J/\ln[(1/2)(-1+\sqrt{5})]
\simeq 2.078|J|$; in the high-temperature phase, $T > T_{p,S}$,
$f_{qn}=(1/2)\ln \lambda_{L,5}$, independent of longitudinal boundary
conditions, but in the low-temperature phase, $T < T_{p,S}$, the expression for
$f_{qn}$ is different for the open and cyclic (equivalently, M\"obius) strips.
There are also other unphysical properties, such as a negative specific heat
and a negative partition function for certain ranges of temperature.  These is
similar to what was found for the analogous strip of the square lattice.

\section{Summary} 

In this paper we have presented exact calculations of the partition function
$Z$ of the $q$-state Potts model and its generalization to real $q$, the Potts 
model, for arbitrary temperature on $n$-vertex strip graphs, of width
$L_y=2$, of the triangular lattice with free, cyclic, and M\"obius longitudinal
boundary conditions. These partition functions are equivalent to Tutte/Whitney
polynomials for these graphs.  The free energy is calculated exactly for the
infinite-length limit of these ladder graphs and the thermodynamics is
discussed.  Considering the full generalization to arbitrary complex $q$ and
temperature, we determine the singular locus ${\cal B}$ in the corresponding
${\mathbb C}^2$ space, arising as the accumulation set of partition function
zeros as $n \to \infty$.  In particular, we study the connection with the $T=0$
limit of the Potts antiferromagnet where ${\cal B}$ reduces to the accumulation
set of chromatic zeros.  Comparisons are made with our previous exact
calculation of Potts model partition functions for the corresponding strips of
the square lattice.  Our present calculations yield, as special cases, several
quantities of graph-theoretic interest, such as the number of spanning trees,
spanning forests, etc., which we record.

Acknowledgment: The research of R. S. was supported in part at Stony Brook by
the U. S. NSF grant PHY-97-22101 and at Brookhaven by the U.S. DOE contract
DE-AC02-98CH10886.\footnote{\footnotesize{Accordingly, the U.S. government
retains a non-exclusive royalty-free license to publish or reproduce the
published form of this contribution or to allow others to do so for
U.S. government purposes.}}

\section{Appendix}

\subsection{General} 

The formulas relating the Potts model partition function $Z(G,q,v)$ and the 
Tutte polynomial $T(G,x,y)$ were given in \cite{a} and hence we shall be brief
here.  The Tutte polynomial of $G$, $T(G,x,y)$, is given by 
\cite{tutte1}-\cite{tutte3}
\beq
T(G,x,y)=\sum_{G^\prime \subseteq G} (x-1)^{k(G^\prime)-k(G)}
(y-1)^{c(G^\prime)}
\label{tuttepol}
\eeq
where $k(G^\prime)$, $e(G^\prime)$, and $n(G^\prime)=n(G)$ denote the number
of components, edges, and vertices of $G^\prime$, and
\beq
c(G^\prime) = e(G^\prime)+k(G^\prime)-n(G^\prime)
\label{ceq}
\eeq
is the number of independent circuits in $G^\prime$. 
As stated in the text, $k(G)=1$ for the graphs of interest here.  Now let
\beq
x=1+\frac{q}{v}
\label{xdef}
\eeq
and
\beq
y=a=v+1
\label{ydef}
\eeq
so that
\beq
q=(x-1)(y-1) \ .  
\label{qxy}
\eeq
Then
\beq
Z(G,q,v)=(x-1)^{k(G)}(y-1)^{n(G)}T(G,x,y) \ .
\label{ztutte}
\eeq

For a planar graph $G$ the Tutte polynomial satisfies the duality relation
\beq
T(G,x,y) = T(G^*,y,x)
\label{tuttedual}
\eeq
where $G^*$ is the (planar) dual to $G$.  As discussed in \cite{a}, 
the Tutte polynomial for recursively defined graphs comprised of $m$ 
repetitions of some subgraph has the form
\beq
T(G_m,x,y) = \sum_{j=1}^{N_\lambda} c_{T,G,j}(\lambda_{T,G,j})^m
\label{tgsum}
\eeq

There are several special cases of the Tutte polynomial that are of interest.
One that we have analyzed in the text and in previous papers is the chromatic
polynomial $P(G,q)$.  This is obtained by setting $y=0$, i.e., $v=-1$, so that
$x=1-q$; the correspondence is $P(G,q) = (-q)^{k(G)}(-1)^{n}T(G,1-q,0)$. 
A second special case is the flow polynomial \cite{bbook,welsh,boll} $F(G,q)$,
obtained by setting $x=0$ and $y=1-q$: 
$F(G,q) = (-1)^{e(G)-n(G)+k(G)}T(G,0,1-q)$  For planar $G$, given the relation
(\ref{tuttedual}), the flow polynomial is, up to a power of $q$, proportional 
to the chromatic polynomial: $F(G,q) \propto P(G^*,q)$.  A third special case 
is the reliability polynomial \cite{welsh}, which will not be considered here. 

\subsection{Triangular Strip with Free Longitudinal Boundary Conditions}

The generating function representation for the Tutte polynomial for the open
strip of the triangular lattice comprised of $m+1$ squares with edges joining
the lower-left to upper right vertices of each square, denoted $S_m$, is
\beq
\Gamma_T(S_m,x,y;z) = \sum_{m=0}^\infty T(S_m,x,y)z^m \ .
\label{gammatfbc}
\eeq
We have
\beq
\Gamma_T(S,x,y;z) = \frac{{\cal N}_T(S,x,y;z)}{{\cal D}_T(S,x,y;z)}
\label{gammas}
\eeq
where
\beq
{\cal N}_T(S,x,y;z)=A_{T,S,0}+A_{T,S,1}z
\label{numts}
\eeq
with
\beq
A_{T,S,0}=x(x+1)^2+2xy+y(y+1)
\label{as0tut}
\eeq
\beq
A_{T,S,1}=-x^3y^2
\label{as1tut}
\eeq
and
\beqs
{\cal D}_T(S,x,y,z) & = & 1-[(x+1)^2 + y(y+2)]z+x^2y^2z^2 \cr\cr
                    & = & \prod_{j=1}^2 (1-\lambda_{T,S,j}z)
\label{dents}
\eeqs
with
\beq
\lambda_{T,S,(1,2)} = \frac{1}{2}\biggl [ (x+1)^2+y(y+2) \pm 
(1+x+y)\sqrt{R_T} \ \biggr ]
\label{lamstut12}
\eeq
where
\beq
R_T=(x+1)^2+(y+1)^2-1-2xy \ . 
\label{rt12}
\eeq

The corresponding closed-form expression is given by the general formula 
\cite{a} 
\beq
T(S_m,x,y)=\biggl [
\frac{A_{T,S,0}\lambda_{T,S,1}+A_{T,S,1}}{\lambda_{T,S,1}-\lambda_{T,S,2}}
\biggr ] (\lambda_{T,S,1})^m + \biggl [
\frac{A_{T,S,0}\lambda_{T,S,2}+A_{T,S,1}}
{\lambda_{T,S,2}-\lambda_{T,S,1}} \biggr ] (\lambda_{T,S,2})^m \ .
\label{tssumform}
\eeq
It is easily checked that this is a symmetric
function of the $\lambda_{S,j}$, $j=1,2$.

\subsection{Cyclic and M\"obius Strips of the Triangular Lattice}

We have 
\beq
T(L_m,x,y) = \sum_{j=1}^6 c_{T,L,j}(\lambda_{T,L,j})^m
\label{tlxy}
\eeq
and
\beq
T(ML_m,x,y) = \sum_{j=1}^6 c_{T,ML,j}(\lambda_{T,ML,j})^m
\label{tmbxy}
\eeq
It is convenient to extract a common factor from the coefficients:
\beq
c_{T,G,j} \equiv \frac{\bar c_{T,G,j}}{x-1} \ , \quad G = L, ML \ .
\label{cbar}
\eeq
Of course, although the individual terms contributing
to the Tutte polynomial are thus rational functions of $x$ rather than
polynomials in $x$, the full Tutte polynomial is a polynomial
in both $x$ and $y$.  We have
\beq
\lambda_{T,ML,j}=\lambda_{T,L,j} \ , \quad j=1,...,6
\label{lamtutlmb}
\eeq
\beq
\lambda_{T,L,1} = 1
\label{lam1tut}
\eeq
and
\beq
\lambda_{T,5}=\lambda_{T,S,1} \ , \quad \lambda_{T,6}=\lambda_{T,S,2}
\label{lam56tut}
\eeq
where $\lambda_{T,S,1}$ and $\lambda_{T,S,2}$ were given in
eq. (\ref{lamstut12}) with (\ref{rt12}). 
The $\lambda_{T,L,j}$, $j=2,3,4$, are solutions of the cubic equation
\beq
\xi^3-(3+2x+2y+y^2)\xi^2+[(x+1)^2+2xy(y+1)]\xi-x^2y^2=0
\label{cubictut}
\eeq
We label these $\lambda_{T,L,j}$'s in a manner corresponding to the
$\lambda_{L,j}$'s occurring in the Potts model partition function, so that 
$\lambda_{T,L,5}$ is dominant in the region in $(x,y)$ equivalent to the PM
phase, and so forth for the others. We note that 
\beq
\lambda_{T,L,2}\lambda_{T,L,3}\lambda_{T,L,4}=\lambda_{T,L,5}\lambda_{T,L,6} 
= x^2y^2 \ .
\label{lamprod}
\eeq

The coefficients for the cyclic strip are 
\beq
\bar c_{T,L,1} = [(x-1)(y-1)]^2-3(x-1)(y-1)+1
\label{c1tutt}
\eeq
\beq
\bar c_{T,L,j} = xy-x-y \quad {\rm for} \quad j=2,3,4
\label{c234tutt}
\eeq
and
\beq
\bar c_{T,L,j} = 1 \quad {\rm for} \quad j=5,6 \ . 
\label{c56tutt}
\eeq
These are symmetric under interchange of $x \leftrightarrow y$, which is a
consequence of the fact that the $\bar c_{L,j}$ are functions only of $q$, 
which, by eq. (\ref{qxy}) is a symmetric function of $x$ and $y$.
For the M\"obius strip, we find 
\beq
\bar c_{T,ML,1}=-1
\label{c1mbtutt}
\eeq
and
\beq
\bar c_{T,ML,5}= \bar c_{T,ML,6}=1 \ .
\label{c56mbtutt}
\eeq
The $c_{T,ML,j}$, $j=2,3,4$ can be calculated from the generating function
using the formula given in \cite{a}.  Their form is more complicated, as is
evident from the results given in \cite{wcy} for the corresponding chromatic
polynomial for this case.  We write the generating function as 
\beq
\Gamma_T(ML,x,y,z) = \sum_{m=1}^\infty T(ML_m,x,y)z^{m-1} \ .
\label{gammatmob}
\eeq
with 
\beq
\Gamma_T(ML,x,y,z) = \frac{ {\cal N}_T(ML,x,y,z)}{{\cal D}_T(ML,x,y,z)}
\label{gammatcalc}
\eeq
where
\beq
{\cal D}_T(ML,x,y,z) = {\cal D}_T(L,x,y,z)=\prod_{j=1}^6 (1-\lambda_{L,j}z)
\label{dst}
\eeq
and
\beq
{\cal N}_T(ML,x,y,z)=\sum_{j=0}^5A_{ML,j}z^j
\label{numgammat}
\eeq
with
\beq
A_{ML,0}=y(x+y+y^2)
\label{a0ml}
\eeq

\beq
A_{ML,1} = yx-4y^2x-4y^3-3y^4-5y^3x-y^5-2x^2y-x^3y-x^2y^2-x^2y^3
     +2y+2x+3x^2+x^3
\label{a1ml}
\eeq

\beqs
A_{ML,2} & = & y(-2x^2y+4yx+3x+3y+x^2y^3-2x^3y+6x^2y^2+2xy^4 \cr\cr
& + & 4y^3x+y^4x^2+3x^3y^2+5y^2+3y^3+y^4+6y^2x+2x^2)
\label{a2ml}
\eeqs

\beqs
A_{ML,3} & = & -y(-2x^2y-3yx-x-y-x^4y+3x^2y^3-2x^3y+8x^2y^2+2xy^4 \cr\cr
& + & 2x^3y^4-x^3y^3+5y^3x-x^3+x^4y^2+2y^4x^2+4x^3y^2-2y^2 \cr\cr
& - & y^3+3y^2x-2x^2)
\label{a3ml}
\eeqs

\beq
A_{ML,4} = x^2y^2(-1+x^2y^3-x^2y^2-x^2+y^2-2x+y+2y^3x+x^2y+y^3+yx)
\label{aml4}
\eeq

and

\beq
A_{ML,5} = -y^4x^4(y-1) \ . 
\label{a5ml}
\eeq

\subsection{Special Values of Tutte Polynomials for Strips of the Triangular
Lattice}

For a given graph $G=(V,E)$, at certain special values of the arguments $x$ and
$y$, the Tutte polynomial $T(G,x,y)$ yields quantities of basic graph-theoretic
interest \cite{tutte3}-\cite{boll}.  We recall some definitions: a spanning
subgraph was defined at the beginning of the paper; a tree is a 
connected graph with no cycles; a forest is a graph containing one or 
more trees; and a spanning tree is a spanning subgraph that is a tree.  We 
recall that the graphs $G$ that we consider are connected.  Then the number 
of spanning trees of $G$, $N_{ST}(G)$, is
\beq
N_{ST}(G)=T(G,1,1) \ ,
\label{t11}
\eeq
the number of spanning forests of $G$, $N_{SF}(G)$, is
\beq
N_{SF}(G)=T(G,2,1) \ ,
\label{t21}
\eeq
the number of connected spanning subgraphs of $G$, $N_{CSSG}(G)$, is
\beq
N_{CSSG}(G)=T(G,1,2) \ ,
\label{T12}
\eeq
and the number of spanning subgraphs of $G$, $N_{SSG}(G)$, is
\beq
N_{SSG}(G)=T(G,2,2) \ .
\label{t22}
\eeq
From the duality relation (\ref{tuttedual}), one has, for planar graphs $G$ and
their planar duals $D^*$, 
\beq
N_{ST}(G)=N_{ST}(G^*) \ , \quad N_{SSG}(G)=N_{SSG}(G^*)
\label{t11t11}
\eeq
and
\beq
N_{SF}(G)=N_{CSSG}(G^*) \ . 
\label{t21t12}
\eeq

 From our calculations of Tutte polynomials, we find that
\beq
N_{ST}(S_m)=N_{ST}(S_m^*) = \Biggl (4+\frac{9\sqrt{5}}{5} \ \Biggr )
\Biggl (\frac{7+3\sqrt{5} \ }{2} \Biggr )^m +
\Biggl (4-\frac{9\sqrt{5}}{5} \ \Biggr )
\Biggl (\frac{7-3\sqrt{5} \ }{2} \Biggr )^m 
\label{t11sm}
\eeq

\beq
N_{SF}(S_m)=N_{CSSG}(S_m^*) = \Bigl ( 12+\frac{17\sqrt{2}}{2} \ \Bigr )
\Bigl [2(3+2\sqrt{2} \ ) \Bigr ]^m +
\Bigl ( 12-\frac{17\sqrt{2}}{2} \ \Bigr )
\Bigl [ 2(3-2\sqrt{2}) \Bigr ]^m 
\label{t21sm}
\eeq

\beq
N_{CSSG}(S_m)=N_{SF}(S_m^*) = (7+5\sqrt{2} \ )
\Bigl [2(3+2\sqrt{2} \ ) \Bigr ]^m + 
(7-5\sqrt{2} \ )\Bigl [2(3-2\sqrt{2} \ ) \Bigr ]^m 
\label{t12sm}
\eeq

\beq
N_{SSG}(S_m)=N_{SSG}(S_m^*) = 2^{4m+5} \ .
\label{t22sm}
\eeq

For the cyclic $L_y=2$ strip of the triangular lattice, $L_m$, we first note 
that for $m \ge 3$, $L_m$ is a (proper) graph, but for $m=1$ and $m=2$, $L_m$
is not a proper graph, but instead, is a multigraph, with multiple edges (and,
for $m=1$, loops).  The following formulas apply for all $m \ge 1$: 
\beq
N_{ST}(L_m)=N_{ST}(L_m^*)=
\frac{2m}{5}\Biggl [\Biggl (\frac{7+3\sqrt{5} \ }{2} \Biggr )^m
+ \Biggl (\frac{7-3\sqrt{5} \ }{2} \Biggr )^m - 2 \Biggr ] 
\label{t11lm}
\eeq
\beq
N_{SF}(L_m)=N_{CSSG}(L_m^*)=
1-\sum_{j=2,3,4} (\lambda_{T,j})^m + \Bigl [ 2(3+2\sqrt{2} \ )
\Bigr ]^m + \Bigl [ 2(3-2\sqrt{2} \ ) \Bigr ]^m \ . 
\label{nsflm}
\eeq
where $\lambda_{T,L,j}$, $j=2,3,4$ are the roots of eq. (\ref{cubictut}) 
for $x=2,y=1$, viz., $\xi^3-10\xi^2+17\xi-4=0$: 
\beq
\lambda_{T,L,2}(x=2,y=1)=0.280176...
\label{lam2tut_21}
\eeq
\beq
\lambda_{T,L,3}(x=2,y=1)=1.803442...
\label{lam3tut_21}
\eeq
\beq
\lambda_{T,L,4}(x=2,y=1)=7.916382...
\label{lam4tut_21}
\eeq
We also calculate 
\beqs
N_{CSSG}(L_m)=N_{SF}(L_m^*) & = & 
-(2+ \frac{6m}{7}) + \biggl [ 1 + m\frac{(3-\sqrt{2} \ )}{7} 
\biggr ] \biggl [ 2(3+2\sqrt{2} \ ) \biggr ]^m \cr\cr
& + & \biggl [ 1 + m\frac{(3+\sqrt{2} \ )}{7} \ \biggr ]
\biggl [ 2(3-2\sqrt{2} \ ) \biggr ]^m
\label{ncssglm}
\eeqs

\beq
N_{SSG}(L_m)=N_{SSG}(L_m^*)=2^{4m} \ . 
\label{nssglm}
\eeq

Since $T(G_m,x,y)$ grows exponentially as $m
\to \infty$ for the families $G_m=S_m$ and $L_m$ for $(x,y)=(1,1)$,
(2,1), (1,2), and (2,2), one defines the corresponding constants
\beq
z_{set}(\{G\}) = \lim_{n(G) \to \infty} n(G)^{-1} \ln N_{set}(G) \ , \quad
set = ST, \ SF, \ CSSG, \ SSG
\label{zset}
\eeq
where, as above, the symbol $\{G\}$ denotes the limit of the graph family $G$
as $n(G) \to \infty$ (and the $z$ here should not be confused with the
auxiliary expansion
variable in the generating function (\ref{gammatfbc}) or the Potts partition
function $Z(G,q,v)$.)  General inequalities for these were given in \cite{a}.
Our results yield 
\beq
z_{ST}(\{G\}) = \frac{1}{2}\ln \Biggl ( \frac{7+3\sqrt{5}}{2} \ \Biggr ) 
\simeq 0.962424 \quad {\rm for} \quad G = S, L, ML
\label{zst}
\eeq
\beq z_{SF}(\{G\}) = z_{CSSG}(\{G\}) = 
\frac{1}{2}\ln[2(3+2\sqrt{2})] \simeq 1.22795
\quad {\rm for} \quad G=S,L,ML
\label{zsfzcssg}
\eeq
and 
\beq
z_{SSG}(\{G\}) = 2\ln 2 \simeq 1.38629 \quad {\rm for} \quad G=S,L,ML
\label{zssg}
\eeq
The cyclic $L_y=2$ strip of the triangular lattice is a $\kappa$-regular graph
\footnote{\footnotesize{A $\kappa$-regular graph is a graph in which all
of the vertices have the same degree, $\kappa$. }} with $\kappa=4$.  It is 
therefore of
interest to compare the value of $z_{ST}$ that we have obtained for the $n \to
\infty$ limit of this family with an upper bound for $\kappa$-regular graphs 
with $\kappa \ge 3$ \cite{mckay,cy} , viz. 
\beq
z_{ST}(\{G\}) \le z_{ST,u.b.}(\{G\}) \ , \quad {\rm where} \quad 
z_{ST,u.b.}(\{G\}) = \ln \Biggl [ \frac{(\kappa-1)^{\kappa-1}}
{[\kappa(\kappa-2)]^{(\kappa/2)-1}} 
\Biggr ] 
\label{zmckay}
\eeq
For this purpose we define $r$ as the ratio of the left- to the right-hand side
of eq. (\ref{zmckay}).  We have
\beq
r_{ST}(\{L\}) = \frac{\frac{1}{2}\ln \Biggl ( \frac{7+3\sqrt{5} }{2} \ \Biggr )
}{3\ln \biggl ( \frac{3}{2} \biggr )} \simeq 0.791
\label{rst}
\eeq
Another comparison of interest is the ratio of $z_{ST}$ for these $L_y=2$
strips with $z_{ST}$ for the full 2D triangular lattice, which has the value 
\cite{wu77}
\beq
z_{ST,tri}=2z_{hc} = \frac{3\sqrt{3}}{\pi}(1 - 5^{-2} + 7^{-2} - 11^{-2} +
13^{-2} - ...) = 1.615329736...
\label{ztrival}
\eeq
\beq
\frac{z_{ST}(\{L\})}{z_{ST}(tri)} \simeq 0.596
\label{zstratio}
\eeq
Finally, we may also compare the $z$ values above with those for the
infinite-length limit of the $L_y=2$ strip of the square lattice (with free
transverse boundary conditions).  This is \cite{fhspan,wust,a} 
\beq
z_{ST} = \frac{1}{2}\ln(2+\sqrt{3} \ ) \simeq 0.658479
\label{zstsq}
\eeq
Further, the calculation of the Tutte polynomial for the open and/or cyclic
$L_y=2$ strip of the square lattice \cite{a} yields 
\beq
z_{SF} = \frac{1}{2}\ln[2(2+\sqrt{3} \ )] \simeq 1.00505
\label{zsfsq}
\eeq
\beq
z_{CSSG} = \frac{1}{2}\ln \Bigl ( \frac{5+\sqrt{17} \ }{2} \Bigr ) \simeq 
0.758832 
\label{zcssgsq}
\eeq
\beq
z_{SSG} = \frac{3}{2}\ln 2 = 1.03972
\label{zssgsq}
\eeq
Thus, one observes that each of these four quantities is greater for the
(infinite-length limit of the) $L_y=2$ strip of the triangular lattice than for
the (same limit of the) $L_y=2$ strip of the square lattice.  For $z_{ST}$,
this follows from the fact that one can obtain the strip of the triangular
lattice from that for the square lattice by uniformly adding a diagonal bond to
each of the squares \cite{stw}.  The inequalities for the other
quantities $z_{SF}$, $z_{CSSG}$, and $z_{SSG}$ are presumably a consequence of
this fact also. 

\subsection{Tutte Polynomials for Dual Graphs}

Since the Tutte polynomial satisfies the duality relation (\ref{tuttedual}) for
a planar graph $G$, our calculations of the Tutte polynomials $T(G_m,x,y)$ for
the open and cyclic $L_y=2$ strips of the triangular lattice, $S_m$ and $L_m$,
also yield the corresponding Tutte polynomials for the duals of these
graphs. For $m \ge 3$, the dual of $L_m$ is a graph with $n=2m+2$ vertices
comprised of the circuit graph $C_{2m}$ together with two additional vertices,
which we denote $v_e$ and $v_o$ such that, if we label the vertices on $C_{2m}$
as $v_j$, $j=1,..,2m$, then each of the $v_j$ with $j$ even is connected by an
edges to $v_e$ and each of the $v_j$ with $j$ odd is connected to $v_o$.  We
shall denote this graph as an ``alternating wheel'', $A_{2m+2}$.  Then
\beq
T(A_{2m+2},x,y) = T(L_m,y,x) \ , \quad {\rm for} \quad m \ge 3
\label{wheeltutte}
\eeq
The cases $m=1$ and $m=2$ can be dealt with in a similar manner. 

The dual of the open strip graph $S_m$ is the multigraph formed by a line of
$2(m+1)$ vertices, of which the interior $2m$ vertices are connected by single
edges to a single additional vertex, and the two end vertices on the line are
connected to this additional vertex by double edges.  We have 
$T((S_m)^*,x,y)=T(S_m,y,x)$.

\vfill
\eject
\end{document}